\documentclass[preprint,12pt]{elsarticle}

\usepackage{lineno}
\usepackage{natbib}
\usepackage{geometry}
\usepackage{fleqn}
\usepackage{graphicx}
\usepackage{newtxtext,newtxmath}
\usepackage{hyperref}
\usepackage{booktabs}
\usepackage{bm}
\usepackage{amsmath}
\usepackage{enumitem}
\usepackage[justification=centering]{caption}
\modulolinenumbers[5]

\usepackage{multirow}
\usepackage{multicol}
\usepackage{subfigure}

\graphicspath{{./figures/}}
\journal{Elsevier} 

\begin{document}

\begin{frontmatter}
	
\title{\large{Towards Improving the Predictive Capability of Computer Simulations by Integrating Inverse Uncertainty Quantification and Quantitative Validation with Bayesian Hypothesis Testing}}

\author[NCSU]{Ziyu Xie}

\author[NCSU]{Farah Alsafadi}

\author[NCSU]{Xu Wu\corref{mycorrespondingauthor}}
\cortext[mycorrespondingauthor]{Corresponding author}
\ead{xwu27@ncsu.edu}

\address[NCSU]{Department of Nuclear Engineering, North Carolina State University    \\ 
	2500 Stinson Drive, Raleigh, NC 27695 \\}

\begin{abstract}
	The Best Estimate plus Uncertainty (BEPU) approach for nuclear systems modeling and simulation requires that the prediction uncertainty must be quantified in order to prove that the investigated design stays within acceptance criteria. A rigorous Uncertainty Quantification (UQ) process should simultaneously consider multiple sources of quantifiable uncertainties: (1) \textit{parameter uncertainty} due to randomness or lack of knowledge; (2) \textit{experimental uncertainty} due to measurement noise; (3) \textit{model uncertainty} caused by missing/incomplete physics and numerical approximation errors, and (4) \textit{code uncertainty} when surrogate models are used. In this paper, we propose a comprehensive framework to integrate results from inverse UQ and quantitative validation to provide robust predictions so that all these sources of uncertainties can be taken into consideration.
	
	Inverse UQ quantifies the parameter uncertainties based on experimental data while taking into account uncertainties from model, code and measurement. In the validation step, we use a quantitative validation metric based on Bayesian hypothesis testing. The resulting metric, called the Bayes factor, is then used to form weighting factors to combine the prior and posterior knowledge of the parameter uncertainties in a Bayesian model averaging process. In this way, model predictions will be able to integrate the results from inverse UQ and validation to account for all available sources of uncertainties. This framework is a step towards addressing the ANS Nuclear Grand Challenge on ``Simulation/Experimentation'' by bridging the gap between models and data.
\end{abstract}

	
\end{frontmatter}


\section{Introduction}

The shift towards multi-physics modeling and simulation (M\&S) has been recently strongly emphasized by the U.S. Department of Energy (DOE), national laboratories, nuclear industry and the U.S. Nuclear Regulatory Commission (NRC). In June 2017, the American Nuclear Society (ANS) announced nine Nuclear Grand Challenges\footnote{\url{https://www.ans.org/challenges/}} that need to be addressed to advance the benefits of nuclear science and technology for future generations. One of them is to \textit{Accelerate Utilization of Simulation and Experimentation}\footnote{\url{https://www.ans.org/challenges/simulation/}}, which aims at integrating M\&S and experimentation to improve predictive simulation capabilities that are necessary to transition nuclear energy system design and licensing from reliance on experiments to reliance on M\&S.

Although the M\&S of nuclear reactors has made tremendous progress, there are always discrepancies between computer designed systems and real world manufactured ones. As a consequence, uncertainties must be quantified along with simulation to facilitate optimal design and decision making, ensure robustness, performance and safety margins. Uncertainty Quantification (UQ) \cite{smith2013uncertainty} is the process to quantify the uncertainties in Quantity-of-Interest (QoIs) by propagating the uncertainties in input parameters through the computer model. The Best Estimate plus Uncertainty (BEPU) approach \cite{wilson2013historical} \cite{rohatgi2020historical} for nuclear systems M\&S requires that the prediction uncertainty must be quantified in order to prove that the investigated designs stay within acceptance criteria.

A rigorous UQ process should simultaneously consider multiple sources of quantifiable uncertainties. Uncertainties in M\&S can be classified as aleatory (irreducible, caused by inherent variation) and epistemic (reducible, caused by lack of knowledge). The four major sources of quantifiable uncertainties in M\&S are: 
\begin{enumerate}[label=(\arabic*)]
	\setlength{\itemsep}{0.1pt}
    \item \textit{Parameter uncertainty}, due to ignorance in the exact values of input parameters (epistemic) or randomness (aleatory). They are associated with design variables or calibration parameters \cite{wu2018inverse-part1}. We assume that uncertainties from design variables are reported in experiments. The uncertainties from calibration parameters can be inferred based on experimental data.
    
    \item \textit{Model uncertainty}, due to missing/inaccurate physics phenomena incorporated in the models, as well as numerical approximation errors. It is also called model discrepancy or model bias \cite{kennedy2001bayesian}. Note that sometimes numerical approximation errors are referred to as a separate type called \textit{numerical uncertainty}. But generally it is difficult to separate it from model uncertainty, and they can be treated in a similar way, so we consider numerical uncertainty as a part of model uncertainty.
    
    \item \textit{Experiment uncertainty}, due to measurement noise/error. It is usually reported with experimental data since the error rates for most instrumentation are known.
    
    \item \textit{Code uncertainty}, due to the emulation of computationally prohibitive codes with surrogate models (also called metamodels).
\end{enumerate}

Failing to account for any of these uncertainties will result in biased predictions. The concept of UQ in the nuclear community generally means forward UQ, in which the information flow is from the inputs to the QoIs. Forward UQ requires knowledge in the input uncertainties, such as the statistical moments (e.g., mean and variance), probability density functions (PDFs), upper and lower bounds, which were usually specified by ``expert opinion'' or ``user self-evaluation''. Inverse UQ, which is the process to inversely quantify the input uncertainties based on experimental data, has recently become popular in the nuclear community. Inverse UQ seeks statistical descriptions of the uncertain input parameters that are consistent with the observation data. In a recent review paper \cite{wu2021comprehensive}, Wu et al. compared and evaluated twelve inverse UQ methods that that have been used on the physical models in system thermal-hydraulics codes. Eight metrics were used to to evaluate an inverse UQ method, including solidity, complexity, accessibility, independence, flexibility, comprehensiveness, transparency, and tractability.

In this paper, \textit{we propose a comprehensive framework to integrate results from inverse UQ and quantitative validation to improve M\&S predictive capability by accounting for all major sources of uncertainties. This framework is a step towards addressing the ANS Nuclear Grand Challenge on ``Simulation/Experimentation'' by bridging the gap between models and data}. In the inverse UQ step, parameter uncertainties are quantified given calibration data. The inverse UQ process uses the modular Bayesian approach (MBA) developed in \cite{wu2018inverse-part1} \cite{wu2018inverse-part2}, which can take into account all four sources of uncertainties simultaneously. In the validation step, the inversely quantified parameter uncertainties are propagated through the computer code to produce QoI predictions with uncertainties. We use a quantitative validation metric based on Bayesian hypothesis testing. The resulting metric, called the Bayes factor (BF) \cite{kass1995bayes} \cite{berger1996intrinsic} \cite{rouder2012default}, is then used to form weighting factors to combine the prior and posterior knowledge of the parameter uncertainties in a Bayesian model averaging (BMA) \cite{hoeting1999bayesian} \cite{wasserman2000bayesian} \cite{raftery2005using} process. In the prediction domain where there is no experimental data available and one has to rely on M\&S to learn about the reality, the predictions will be made by averaging the computational results from the prior and posterior models, weighted by the probabilities that each model is more accurate. Such integration will be able to combine information from prior knowledge (expert judgment), inverse UQ (calibration) and quantitative validation.

There are a few previous works that also focused on a holistic uncertainty analysis of M\&S predictions. Roy and Oberkampf \cite{roy2011comprehensive} developed a framework for verification, validation, and uncertainty quantification (VVUQ) to estimate the predictive uncertainty. This framework considers both aleatory and epistemic uncertainties and characterizes them as random variables and interval-valued quantities, respectively. Another notable feature is that it includes the contribution from numerical approximation errors and model form uncertainty, which are estimated using verification and validation extrapolation, respectively. However, this VVUQ framework does not involve calibration or inverse UQ, the input uncertainty information is therefore obtained from expert opinion. Forward UQ from input to QoIs becomes more complicated by treating inputs as aleatory and epistemic, as ``double looping'' (also called nested sampling) is required.

Li et al. \cite{li2016integrating} developed an integrated Bayesian calibration, bias correction, and machine learning (ML) approach. The model predictions were made using the ML model and corrected with the bias model. This approach allowed the quantification of multiple sources of uncertainty and errors in the predictions, as well as improved model performance in untested domains by combining multiple sources of information. Jiang et al. \cite{jiang2020sequential} proposed a sequential model calibration and validation (SeCAV) framework that can improve the efficacy of both model parameter calibration and model bias correction for the purpose of quantification and reduction of uncertainty. SeCAV implements model validation and Bayesian calibration in a sequential manner. The validation step serves as a filter to select the best experimental data for the calibration step, and provides a confidence probability as a weight factor to update the model parameter uncertainties. Results from the calibration step is then integrated with model bias correction to improve the prediction accuracy.

Verification, calibration, and validation are straightforward for single-level, or single-component systems. However, complex systems usually have multiple models interact with each other, for example, the output of a lower-level model becomes an input to a higher-level model. Propagation of uncertainty through the whole system by integrating verification, calibration, and validation is necessary. Sankararaman et al. \cite{sankararaman2015integration} proposed a Bayesian method to integrate verification, calibration, and validation for complex multi-level engineering systems. They used a Bayesian network to connect the various models, their inputs/outputs, experimental data, and various sources of model error. The Bayesian network can combine the results of verification, calibration, and validation for each individual model using the principles of conditional/total probability, and allowing various uncertainties sources to be propagated to quantify the overall system-level uncertainty. Mullins et al. \cite{mullins2016bayesian} developed a Bayesian methodology to incorporate information at different levels by integrating calibration and validation when making predictions. Particularly, this method included the impact of sparse data in the prediction uncertainty and handled both aleatory and epistemic uncertainties. Li et al. \cite{li2016role} quantified the multi-level system prediction uncertainty by integrating calibration, validation and sensitivity analysis at different levels. A model reliability metric was used for quantitative model validation at each lower level. A relevance study between the lower-level models and system-level response was performed using sensitivity analysis. Results from calibration, validation and sensitivity analysis were combined to obtain the integrated distribution of model parameters, which were propagated through the multi-level system to quantify the overall system prediction uncertainty.

This paper is organized as follows. Section 1 introduces the basic concepts and reviews previous works. Section 2 presents the methodology, which consists of inverse UQ, quantitative validation, and integration of inverse UQ and validation results. Section 3 briefly introduces the experimental data used for the demonstration application, while Section 4 presents the results and discussion. Section 5 concludes the paper.

\section{Methodology}

\begin{figure}[htbp]
	\centering
	\includegraphics[width=0.99\textwidth]{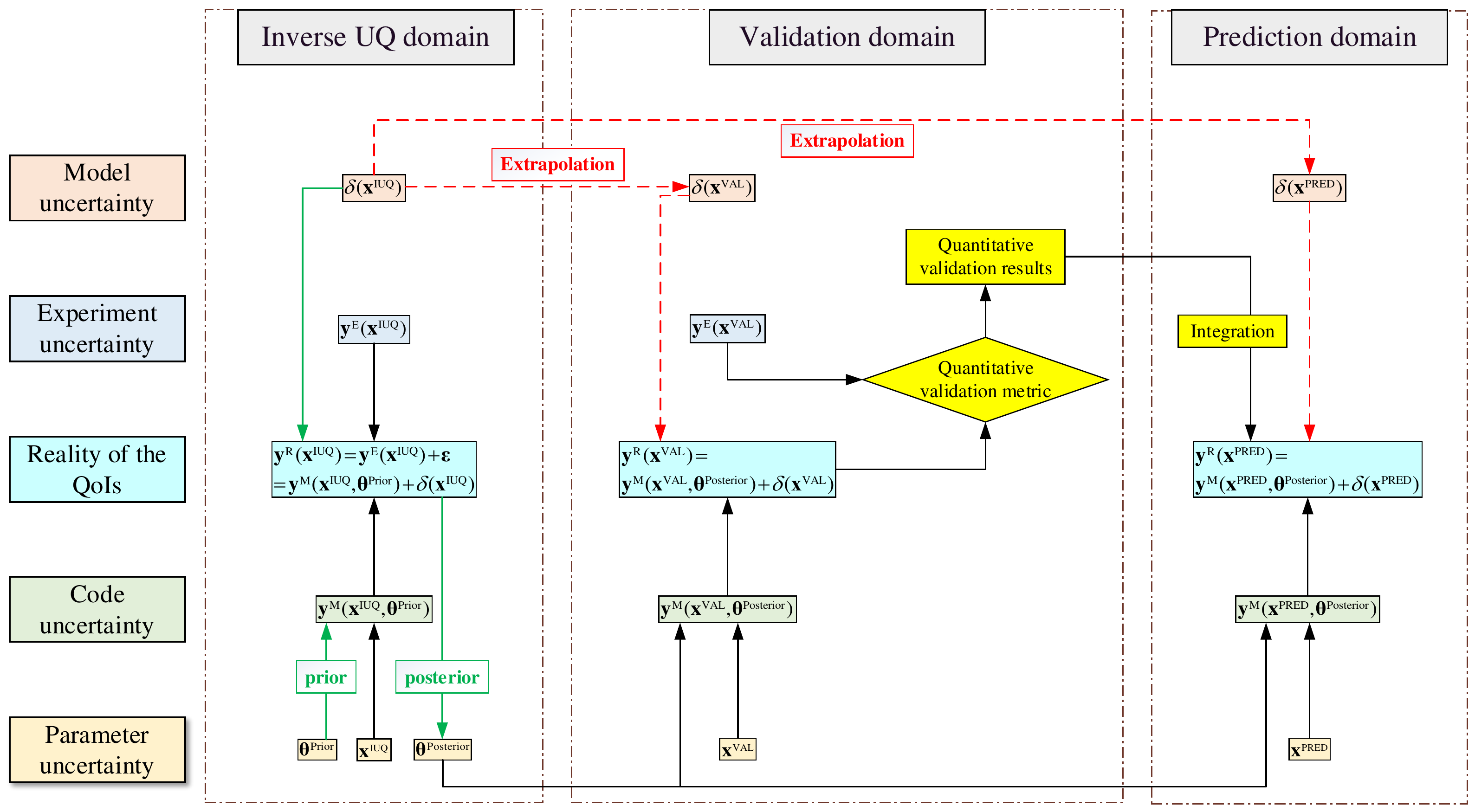}
	\caption[]{The proposed framework to integrate inverse UQ and quantitative validation.}
	\label{figure:fig1-Integrated-framework}
\end{figure}

Figure \ref{figure:fig1-Integrated-framework} shows the proposed framework to enhance the predictive capabilities of computer models by integrating results from inverse UQ and validation within the Bayesian framework. In Figure \ref{figure:fig1-Integrated-framework}, the superscripts ``R'', ``M'' and ``E'' stand for reality, model and experiment respectively, while the superscripts ``IUQ'', ``VAL'' and ``PRED'' represent domains for Inverse UQ, VALidation and PREDiction respectively. The meanings of all the mathematical symbols are presented in Table \ref{table:table1-Symbols}.

The framework consists of three major steps. The first step is \textit{inverse UQ} \cite{wu2018inverse-part1}, or (statistical/Bayesian) calibration. In this step, the unknown parameter uncertainties are inversely quantified while keeping model simulation consistent with calibration data. The model uncertainty can also be quantified. The second step is quantitative \textit{validation}, which is the step to quantitatively evaluate the accuracy of calibrated models in representing the real world through comparison with the validation data. The third step is \textit{prediction}, which combines results from inverse UQ and validation to make predictions in domains where there is no data and one has to rely on simulation to learn about the reality.

\begingroup
\renewcommand{\arraystretch}{1.2}
\begin{table}[!ht]
	\footnotesize
	\centering
	\captionsetup{justification=centering}
	\caption{Definitions of symbols used in the integrated framework.}
	\label{table:table1-Symbols}
	\begin{tabular}{c c}
		\toprule
		Symbol  &  Description  \\ 
		\midrule
		$\mathbf{x}$                      &  design variables  \\
		$\bm{\theta}$                     &  calibration parameters  \\
		$\mathbf{y}$                      &  responses, or QoIs  \\
		$\mathbf{y}^{\text{M}}$           &  QoIs from model simulation  \\
		$\mathbf{y}^{\text{R}}$           &  QoIs' unknown real values  \\
		$\mathbf{y}^{\text{E}}$           &  QoIs from experiment  \\
		$\mathbf{x}^{\text{IUQ}}$         &  inverse UQ domain       \\
		$\mathbf{x}^{\text{VAL}}$         &  validation domain       \\
		$\mathbf{x}^{\text{PRED}}$        &  prediction domain       \\
		$\bm{\theta}^{*}$                 &  true but unknown values of $\bm{\theta}$  \\
		$\bm{\theta}^{\text{Prior}}$      &  prior distributions of $\bm{\theta}$  \\
		$\bm{\theta}^{\text{Posterior}}$  &  posterior distributions of $\bm{\theta}$  \\
		$\bm{\epsilon}$                   &  measurement noise   \\
		$\delta(\mathbf{x})$              &  model discrepancy/bias     \\
		$\bm{\Sigma}$                     &  total uncertainty of the posterior PDF  \\
		$\bm{\Sigma}_{\text{exp}}$        &  experimental uncertainty  \\
		$\bm{\Sigma}_{\text{bias}}$       &  model uncertainty  \\
		$\bm{\Sigma}_{\text{code}}$       &  code uncertainty  \\
		\bottomrule
	\end{tabular}
\end{table}
\endgroup

The framework in Figure \ref{figure:fig1-Integrated-framework} is comprehensive in the sense that it incorporates uncertainties from four major sources, which are listed in four of the five rows of Figure \ref{figure:fig1-Integrated-framework} (the center row represents the unknown reality). They are uncertainties from parameter, experiment, model, and code. These four sources include the most, if not all, of the quantifiable uncertainties in M\&S that should be considered when making predictions. In the following three subsections, the theories for inverse UQ, validation and prediction will be presented.

\subsection{Bayesian inverse UQ}

The process of propagating uncertainties from input parameters to QoIs is widely known as forward UQ. Forward UQ requires input uncertainty information which has been usually specified using ``expert opinion'' or ``user self-evaluation''. Such ad hoc specification lacks mathematical rigor and can be subjective, leading to biased forward UQ results. Inverse UQ is the process to inversely quantify the input uncertainties while keeping model outputs consistent with measurement data. An innovative inverse UQ method based on Bayesian inference was developed in \cite{wu2018inverse-part1} and successfully applied to system thermal-hydraulics code TRACE \cite{wu2018inverse-part2} and fuel performance code BISON \cite{wu2018Kriging}. The resulting posterior distributions can effectively represent the input parameter uncertainties that are consistent with data.

Consider a general computer model $\mathbf{y}^{\text{M}} = \mathbf{y}^{\text{M}} ( \mathbf{x}, \bm{\theta} )$ where $\mathbf{y}^{\text{M}}$ is the model QoI, $\mathbf{x}$ is the vector of design variables, and $\bm{\theta}$ is the vector of calibration parameters. See \cite{wu2018inverse-part1} for a detailed discussion on the differences between design variables and calibration parameters. Given an experimental condition characterized by design variables $\mathbf{x}$, the reality $\mathbf{y}^{\text{R}} (\mathbf{x})$ can be predicted by computer model simulation. Because the computer model is only an approximation of the reality, we have the relation shown in Equation (\ref{equation:IUQ1-Bias}):
\begin{equation}	 \label{equation:IUQ1-Bias}
\mathbf{y}^{\text{R}} (\mathbf{x}) = \mathbf{y}^{\text{M}} \left( \mathbf{x}, \bm{\theta}^{*} \right) + \delta(\mathbf{x})
\end{equation}
where $\bm{\theta}^{*}$ is the ``true'' but unknown values for $\bm{\theta}$. The learning of $\bm{\theta}^{*}$ is the goal of inverse UQ. $\delta(\mathbf{x})$ is the \textit{model bias/discrepancy}, also called \textit{model uncertainty}, \textit{model inadequacy} or \textit{model error} \cite{kennedy2001bayesian}. The model bias is due to incomplete or inaccurate underlying physics, numerical approximation errors, and/or other inaccuracies that would exist even if all the parameters in the computer model were known \cite{arendt2012quantification}. The reality can also be learned by performing experiments. Define experimental observations as $\mathbf{y}^{\text{E}} (\mathbf{x})$, we also have:
\begin{equation}
	\mathbf{y}^{\text{E}} (\mathbf{x}) = \mathbf{y}^{\text{R}} (\mathbf{x}) + \bm{\epsilon}
\end{equation}
where $\bm{\epsilon} \sim \mathcal{N} ( 0, \Sigma_{\text{exp}} )$ represents the \textit{measurement noise/error}. Combining the above two equations:
\begin{equation}	 \label{equation:IUQ2-Model-Update-Eqn}
	\mathbf{y}^{\text{E}} (\mathbf{x}) = \mathbf{y}^{\text{M}} \left( \mathbf{x}, \bm{\theta}^{*} \right) + \delta(\mathbf{x}) + \bm{\epsilon}
\end{equation}

Equation (\ref{equation:IUQ2-Model-Update-Eqn}) is referred to as ``\textit{model updating formulation}'' \cite{kennedy2001bayesian} \cite{arendt2012quantification}, which serves as the starting point of Bayesian inverse UQ. Based on the model updating equation and the Gaussian assumption of the experiment uncertainty, $\bm{\epsilon} = \mathbf{y}^{\text{E}} (\mathbf{x}) - \mathbf{y}^{\text{M}} ( \mathbf{x}, \bm{\theta}^{*} ) - \delta(\mathbf{x})$ follows a multi-dimensional Gaussian distribution. The posterior can be written as:
\begin{multline}	 \label{equation:IUQ3-Posterior}
	p \left( \bm{\theta}^{*} | \mathbf{y}^{\text{E}}, \mathbf{y}^{\text{M}}\right)  \propto  p \left( \bm{\theta}^{*} \right) \cdot p \left( \mathbf{y}^{\text{E}}, \mathbf{y}^{\text{M}} | \bm{\theta}^{*} \right) 	\\
	\propto  p \left( \bm{\theta}^{*} \right) \cdot \frac{1}{\sqrt{|\bm{\Sigma}|}}  \cdot \text{exp} \left[  - \frac{1}{2} \left[ \mathbf{y}^{\text{E}} - \mathbf{y}^{\text{M}} - \delta \right]^\top \bm{\Sigma}^{-1} \left[ \mathbf{y}^{\text{E}} - \mathbf{y}^{\text{M}} - \delta \right] \right]
\end{multline}
where $p ( \bm{\theta}^{*} )$ is the prior distribution and $p ( \mathbf{y}^{\text{E}}, \mathbf{y}^{\text{M}} | \bm{\theta}^{*} )$ is the likelihood function. Prior and posterior probabilities represent degrees of belief about possible values of $\bm{\theta}^{*}$, before and after observing the experimental data. $\bm{\Sigma}$ is the covariance of the posterior PDF that consists of three parts:
\begin{equation}
	\bm{\Sigma} = \bm{\Sigma}_{\text{exp}} + \bm{\Sigma}_{\text{bias}} + \bm{\Sigma}_{\text{code}}
\end{equation}

The first part $\bm{\Sigma}_{\text{exp}}$ is the \textit{experimental uncertainty} due to measurement error. The second part $\bm{\Sigma}_{\text{bias}}$ represents the \textit{model uncertainty/bias}. The third term $\bm{\Sigma}_{\text{code}}$ is called \textit{code uncertainty}, or \textit{interpolation uncertainty}, because we do not know the computer code outputs at every input, especially when the code is computationally prohibitive. In this case, a metamodel can be used to reduce the computational cost. Metamodels, also called surrogate models, response surfaces, or emulators, are approximations of the input/output relation of a computer model. Metamodels are built from a limited number of original model runs (training set) and a learning algorithm, and they can be evaluated very quickly. Note that $\bm{\Sigma}_{\text{code}} = \mathbf{0}$ if the computer model is used instead of its surrogate models.

Bayesian inverse UQ uses Markov Chain Monte Carlo (MCMC) sampling \cite{andrieu2008tutorial} \cite{brooks2011handbook} to generate samples that follow a probability density proportional to the posterior PDF in Equation (\ref{equation:IUQ3-Posterior}), without knowing the normalizing constant. MCMC requires a large number of samples to sufficiently explore the posterior space, which is expensive for computationally prohibitive models. Metamodeling techniques, such as Gaussian Process (GP) \cite{santner2003design} is recommended. The greatest advantage of GP is that it directly provides an estimation of the code uncertainty $\bm{\Sigma}_{\text{code}}$.  In this way, once does not become over-confident in the GP surrogate model, due to the code uncertainty term in the variance of the posterior PDF.

The inverse UQ method developed in \cite{wu2018inverse-part1} is called the modular Bayesian approach (MBA). Compared to the traditional full Bayesian approach (FBA) \cite{kennedy2001bayesian}, modularization was introduced to separate various modules in Bayesian inverse UQ to prevent suspect information belonging to one part from overly influencing another part. The MBA method has reduced complexity from reasonable simplification and better convergence for MCMC sampling, compared to FBA. The most significant characteristic of MBA is the simultaneous consideration of all major sources of quantifiable uncertainties in M\&S, i.e., uncertainties from parameter, experiment, model and code.

\subsection{Quantitative validation}

Verification and Validation (V\&V) \cite{oberkampf2010verification} are the primary techniques to build and quantify the confidence in M\&S. Verification is the process of determining that a model implementation accurately represents the developer’s conceptual description of the model and the solution to the model. Validation is the process of determining the degree to which a model is an accurate representation of the real world from the perspective of the intended uses of the model. In this work, the computer models are assumed to be already verified by the model developers. Therefore, the primary focus in this part is validation.

\subsubsection{Problems with graphical validation}

Even though there is a growing interest in performing validation and a sizable literature on this topic among the nuclear community, most of them were confined to \textit{graphical validation}: an approach to find model agreement with data by plotting them together. It was often declared that the model is ``validated'' if the overall visual agreement is satisfactory, for example, if the data points fall on or closely to the diagonal line, or most of the data points stay within the error bars of the simulation results. However, such graphical comparison is inadequate for accessing the accuracy of computer models due to the following reasons: (1) graphical comparison is only qualitative and validation decisions are therefore often subjective; (2) for cases when the model accuracy varies over spatial or temporal ranges, graphical comparison gives very limited quantitative information on such variation; (3) information in the validation data cannot be readily passed to the prediction domain; (4) graphical comparison may not give direct suggestion on which model performs better for model selection. For example, models A and B may perform better on different regions, making it difficult to make a fair selection only based on visual inspection.

In this work we propose to implement a quantitative validation metric to replace graphical comparison for a more rigorous validation process. \textit{Validation metrics} \cite{oberkampf2006measures} are quantitative measures of the level of agreement, or consistency, between computational results and experimental data. Furthermore, we need a validation metric that can be seamlessly incorporated into the proposed framework in Figure \ref{figure:fig1-Integrated-framework}, for example, by providing weight factors to combine the inverse UQ and validation results. We will use Bayesian hypothesis testing \cite{ling2013quantitative} \cite{liu2011toward} as a quantitative validation metric. In \textit{classical hypothesis testing}, one starts with making a null hypothesis that the observation data comes from the model prediction population. Then an unbiased test statistic is constructed and evaluated based on data and model predictions. Finally one decides whether there is enough evidence to reject or not reject this null hypothesis. In \textit{Bayesian hypothesis testing}, prior and posterior distributions of model responses are compared to form a metric called Bayes factor (BF) \cite{kass1995bayes} \cite{berger1996intrinsic} \cite{rouder2012default}, which is the ratio of posterior and prior QoI PDFs based on the support from validation data. The posterior and prior QoI PDFs are obtained by forward propagating the posterior and prior parameter uncertainties through the model, respectively.

\subsubsection{Bayesian hypothesis testing and the Bayes factor}

In hypothesis testing, one has $H_0$ as the \textit{Null Hypothesis (NH)} and $H_1$ as the \textit{Alternative Hypothesis (AH)}. For example, in a validation setting, examples of $H_0$ and $H_1$ can be: (1) a model is valid vs. invalid for the intended use; (2) model A is more accurate vs. model B is more accurate for two competing models; (3) for the same model, calibrated version is more accurate vs. uncalibrated version is more accurate. For notational convenience, define inverse UQ data as $\mathbf{y}^{\text{E}} ( \mathbf{x}^{\text{IUQ}} ) = D^{\text{IUQ}}$. Similarly, define validation data as $\mathbf{y}^{\text{E}} ( \mathbf{x}^{\text{VAL}} ) = D^{\text{VAL}}$. Based on the Bayes’ rule $P(A|B) = P(B|A) \cdot P(A) / P(B)$, we have:
\begin{equation}
	\begin{aligned}
		P \left( H_0 | D^{\text{VAL}} \right) = \frac{P \left( D^{\text{VAL}} | H_0 \right) \cdot P \left( H_0 \right) }{P (D^{\text{VAL}})}		\\
		P \left( H_1 | D^{\text{VAL}} \right) = \frac{P \left( D^{\text{VAL}} | H_1 \right) \cdot P \left( H_1 \right) }{P (D^{\text{VAL}})}
	\end{aligned}
\end{equation}

Here $H_0$ and $H_1$ are hypotheses that \textit{calibrated} and \textit{uncalibrated} models are more accurate, respectively. In other words, we are comparing models whose parameter uncertainties are characterized by \textit{posterior} and \textit{prior distributions}, respectively. Taking the ratio of the above two equations:
\begin{equation}
	\frac{P \left( H_0 | D^{\text{VAL}} \right)}{P \left( H_1 | D^{\text{VAL}} \right)} = \frac{P \left( D^{\text{VAL}} | H_0 \right)}{P \left( D^{\text{VAL}} | H_1 \right)} \cdot \frac{P ( H_0 )}{P ( H_1 )}
\end{equation}

The BF $B$ is defined as the likelihood ratio of $H_0$ and $H_1$. It compares the support for $H_0$ and $H_1$ from the validation data $D^{\text{VAL}}$. If $B>1$, the model for $H_0$ is more accurate (the model corresponds to $H_0$ has more support from $D^{\text{VAL}}$) and vice versa.
\begin{equation}
	B = \frac{P \left( D^{\text{VAL}} | H_0 \right)}{P \left( D^{\text{VAL}} | H_1 \right)}
\end{equation}

\subsubsection{Calculation of the Bayes factor}

Calculation of the BF $B$ is a complicated process but can be solved using Monte Carlo sampling-based numerical integration. Because large number of samples are needed, surrogate models can also be used to reduce the computational cost, similar to inverse UQ.
\begin{equation}	 \label{equation:BF1}
    B = \frac{P \left( D^{\text{VAL}} | H_0 \right)}{P \left( D^{\text{VAL}} | H_1 \right)} 
    = \frac{ \int f \left( D^{\text{VAL}} | \mathbf{y}^{\text{M}}, H_0 \right) \cdot f_{\mathbf{Y}^{\text{M}}} \left( \mathbf{y}^{\text{M}} | H_0 \right) d \mathbf{y}^{\text{M}} }{ \int f \left( D^{\text{VAL}} | \mathbf{y}^{\text{M}}, H_1 \right) \cdot f_{\mathbf{Y}^{\text{M}}} \left( \mathbf{y}^{\text{M}} | H_1 \right) d \mathbf{y}^{\text{M}} } 
\end{equation}

The first part in the integrals of the numerator and denominator, $f ( D^{\text{VAL}} | \mathbf{y}^{\text{M}}, H_0 )$ and $f ( D^{\text{VAL}} | \mathbf{y}^{\text{M}}, H_1 )$, are multivariate Gaussian distributions because of Equation (\ref{equation:IUQ2-Model-Update-Eqn}). The variance of the Gaussian distribution is a combination of model bias $\delta(\mathbf{x})$ and the measurement error $\bm{\epsilon}$. Note that it is unwise to use the model bias $\delta(\mathbf{x})$ learned during inverse UQ, because $\delta(\mathbf{x})$ is dependent on the domain define by the design variable $\mathbf{x}$. Using $\delta(\mathbf{x})$ obtained from the inverse UQ domain $\mathbf{x}^{\text{IUQ}}$ on the validation domain $\mathbf{x}^{\text{VAL}}$ may result in large errors due to extrapolation (red arrows/blocks shown in Figure \ref{figure:fig1-Integrated-framework}). If there is no accurate estimation of $\delta(\mathbf{x}^{\text{VAL}})$ available, it is reasonable to ignore the contribution to the Gaussian variance from model bias.

The second parts, $f_{\mathbf{Y}^{\text{M}}} ( \mathbf{y}^{\text{M}} | H_0 )$ and $f_{\mathbf{Y}^{\text{M}}} ( \mathbf{y}^{\text{M}} | H_1 )$ are the probabilities of model prediction $\mathbf{y}^{\text{M}}$ given the models $H_0$ and $H_1$, respectively. They can also be obtained by numerical integration as shown below:

\begin{equation}
	\begin{aligned}
		f_{\mathbf{Y}^{\text{M}}} \left( \mathbf{y}^{\text{M}} | H_0 \right) &= 
		\int f_{\mathbf{Y}^{\text{M}}} \left( \mathbf{y}^{\text{M}} | \mathbf{x}, \bm{\theta}, H_0 \right) \cdot f_{\mathbf{X}} (\mathbf{x}) \cdot f_{\bm{\Theta}} \left( \bm{\theta} | H_0 \right)  d \mathbf{x} d \bm{\theta}
		\\
		f_{\mathbf{Y}^{\text{M}}} \left( \mathbf{y}^{\text{M}} | H_1 \right) &= 
		\int f_{\mathbf{Y}^{\text{M}}} \left( \mathbf{y}^{\text{M}} | \mathbf{x}, \bm{\theta}, H_1 \right) \cdot f_{\mathbf{X}} (\mathbf{x}) \cdot f_{\bm{\Theta}} \left( \bm{\theta} | H_1 \right)  d \mathbf{x} d \bm{\theta}
	\end{aligned}
\end{equation}
where $f_{\mathbf{Y}^{\text{M}}} ( \mathbf{y}^{\text{M}} | \mathbf{x}, \bm{\theta}, H_i ) (i=0,1)$ denotes the conditional probabilities of $\mathbf{y}^{\text{M}}$ given $\mathbf{x}$ and $\bm{\theta}$ for model $H_i$. $f_{\mathbf{X}} (\mathbf{x})$ is the probability of the design variables $\mathbf{x}$, which are not treated as uncertain in this work, so we have $f_{\mathbf{X}} (\mathbf{x}) = 1$. Note that it is also common to have uncertainties in the design variables in the UQ applications. If so, these uncertainties can be considered easily based on the above equation. Finally, $f_{\bm{\Theta}} ( \bm{\theta} | H_i ) (i=0,1)$ represents the parameter uncertainties for $\bm{\theta}$. Note that $f_{\bm{\Theta}} ( \bm{\theta} | H_0 )$ implies a dependence on $D^{\text{IUQ}}$, i.e., $f_{\bm{\Theta}} ( \bm{\theta} | H_0 ) = f_{\bm{\Theta}} ( \bm{\theta} | H_0, D^{\text{IUQ}} )$ because it represents the posterior distributions of the parameter uncertainties after inverse UQ, while $f_{\bm{\Theta}} ( \bm{\theta} | H_1 )$ denotes the prior distributions based on expert judgment before inverse UQ. Equation (\ref{equation:BF1}) can be formulated as:
\begin{equation}	 \label{equation:BF2}
    B =  \frac{ \int f \left( D^{\text{VAL}} | \mathbf{y}^{\text{M}}, H_0 \right) \cdot \left( \int f_{\mathbf{Y}^{\text{M}}} \left( \mathbf{y}^{\text{M}} | \bm{\theta}, H_0 \right) \cdot f_{\bm{\Theta}} \left( \bm{\theta} | H_0 \right)  d \bm{\theta} \right) d \mathbf{y}^{\text{M}} }{ \int f \left( D^{\text{VAL}} | \mathbf{y}^{\text{M}}, H_1 \right) \cdot \left( \int f_{\mathbf{Y}^{\text{M}}} \left( \mathbf{y}^{\text{M}} | \bm{\theta}, H_1 \right) \cdot f_{\bm{\Theta}} \left( \bm{\theta} | H_1 \right)  d \bm{\theta} \right) d \mathbf{y}^{\text{M}} } 
\end{equation}

The BF can be used as a quantitative validation metric to measure the degree of simulation accuracy instead of graphical comparison. More importantly, the BF will be used to form weight factors in a BMA process in the prediction domain, as shown in the next section. In this way, the information from validation data will be used in the prediction domain, instead of being discarded after graphical validation.

\subsection{Prediction by integrating inverse UQ and validation results}

The quantitative validation metric BF will be used to integrate inverse UQ and validation results. Since we are only dealing with two hypotheses, we have:
\begin{equation}
	\begin{aligned}
		P ( H_0 ) + P ( H_1 ) &= 1   \\
		P ( H_0 | D^{\text{VAL}} ) + P ( H_1 | D^{\text{VAL}} ) &= 1
	\end{aligned}
\end{equation}

Furthermore, we assume no prior preferences for $H_0$ and $H_1$ so they are equally likely, i.e., $P ( H_0 ) = P ( H_1 ) = 0.5$. Then we have:
\begin{equation}
    \frac{P ( H_0 | D^{\text{VAL}} )}{P ( H_1 | D^{\text{VAL}} )}  =  \frac{P ( H_0 | D^{\text{VAL}} )}{1 - P ( H_0 | D^{\text{VAL}} )}  =  B  \cdot \frac{P ( H_0 )}{P ( H_1 )}  =  B
\end{equation}

The posterior probabilities of the two models conditioned on the validation data become:
\begin{equation}	 \label{equation:Prediction1-Weight-Factor}
    \begin{aligned}
        P ( H_0 | D^{\text{VAL}} ) &= \frac{B}{B+1}    \\
        P ( H_1 | D^{\text{VAL}} ) &= \frac{1}{B+1}
    \end{aligned}
\end{equation}

It is shown in Equation (\ref{equation:Prediction1-Weight-Factor}) that the posterior probability (not to be confused with the parameter posterior distributions) of the model corresponds to $H_0$ based on the validation data $D^{\text{VAL}}$ can be expressed as a simple function of the BF. A special case is when the BF $B$ equals 1, i.e., the validation data $D^{\text{VAL}}$ has equal support for $H_0$ and $H_1$, we will have $P ( H_0 | D^{\text{VAL}} ) = P ( H_1 | D^{\text{VAL}} ) = 0.5$. This important result in Equation (\ref{equation:Prediction1-Weight-Factor}) will be used in the integration of inverse UQ and validation results.

\begin{figure}[htbp]
	\centering
	\includegraphics[width=0.99\textwidth]{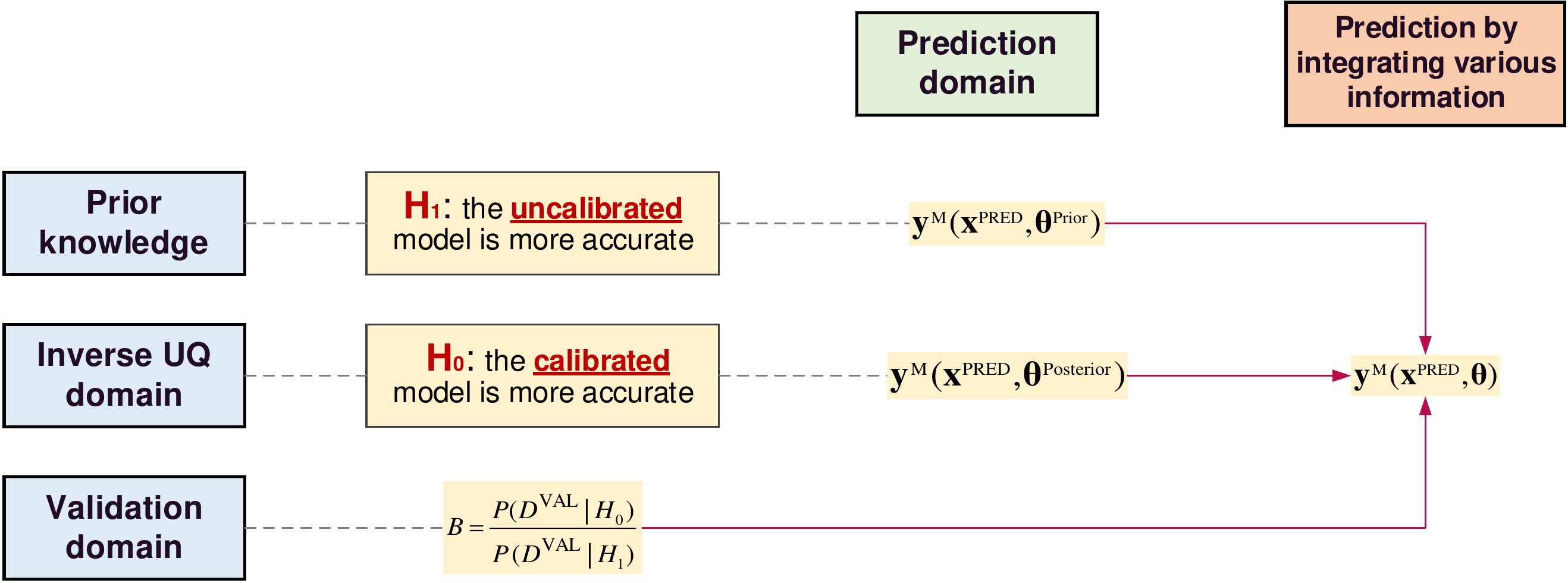}
	\caption[]{Using the Bayes factor to integrate inverse UQ and validation results.}
	\label{figure:fig2-Prediction-Weighted-Average}
\end{figure}

The BF can be used to formulate weighting factors to combine prior (uncalibrated) and posterior (calibrated) models in a BMA process. The prediction based on BMA is formulated by Equation (\ref{equation:Prediction2-Weighted-Average}) and illustrated in Figure \ref{figure:fig2-Prediction-Weighted-Average}.
\begin{equation}
	\begin{aligned}	 \label{equation:Prediction2-Weighted-Average}
		\mathbf{y}^{\text{M}} ( \mathbf{x}^{\text{PRED}}, \bm{\theta} )
		&= P ( H_0 | D ) \cdot \mathbf{y}^{\text{M}} ( \mathbf{x}^{\text{PRED}}, \bm{\theta}^{\text{Posterior}} )  +  P ( H_1 | D ) \cdot \mathbf{y}^{\text{M}} ( \mathbf{x}^{\text{PRED}}, \bm{\theta}^{\text{Prior}} )  	\\
		&= \frac{B}{B+1} \cdot \mathbf{y}^{\text{M}} ( \mathbf{x}^{\text{PRED}}, \bm{\theta}^{\text{Posterior}} )  +  \frac{1}{B+1} \cdot \mathbf{y}^{\text{M}} ( \mathbf{x}^{\text{PRED}}, \bm{\theta}^{\text{Prior}} )
	\end{aligned}
\end{equation}

Equation (\ref{equation:Prediction2-Weighted-Average}) states that in the prediction domain where there is no data, one can make predictions by averaging the computational results from the uncalibrated and calibrated models, weighted by the probabilities that each model is more accurate based on the information from the validation domain. Such integration will be able to combine information from prior knowledge (mainly based on expert opinion), inverse UQ (calibration) and quantitative validation.

\subsection{Summary on the workflow of the integration framework}

After introducing the theories for inverse UQ, quantitative validation, and prediction, the main workflow of the integration framework in Figure \ref{figure:fig1-Integrated-framework} can be summarized as:
\begin{enumerate}[label=(\alph*)]
	\setlength{\itemsep}{0.1pt}
	\item In the inverse UQ domain, given the calibration data $\mathbf{y}^{\text{E}} ( \mathbf{x}^{\text{IUQ}} )$, or $D^{\text{IUQ}}$, the knowledge about the calibration parameters $\bm{\theta}$ is updated from $\bm{\theta}^{\text{Prior}}$ to $\bm{\theta}^{\text{Posterior}}$. The model bias $\delta(\mathbf{x})$ needs to be considered during inverse UQ. If $\delta(\mathbf{x})$ is not considered, Equation (\ref{equation:IUQ1-Bias}) is reduced to $\mathbf{y}^{\text{R}} (\mathbf{x}) = \mathbf{y}^{\text{M}} \left( \mathbf{x}, \bm{\theta}^{*} \right)$. In other words, we are over-confident in the computer model because we treat the model prediction as the reality. This can potentially cause over-fitting in the posterior distributions, which means that the posterior distributions are over-fitted to the calibration data such that when applied to new experiments, the model's prediction accuracy will be low. Model bias $\delta(\mathbf{x})$ can be considered using MBA or FBA, see detailed discussion in \cite{wu2021comprehensive} \cite{wu2018inverse-part1}.
	
	\item In the validation domain, given the validation data $\mathbf{y}^{\text{E}} ( \mathbf{x}^{\text{VAL}} )$, or $D^{\text{VAL}}$, the accuracy of the calibrated model (using posterior uncertainties) can be compared with the uncalibrated model (using prior uncertainties) based on a quantitative validation metric, the BF. If the BF is greater than 1.0, the calibrated model has a better accuracy than the uncalibrated model, and vice versa. Besides providing a more robust validation approach compared to visual inspection, BF can also carry the information in the validation data to the prediction domain.
	
	\item In the prediction domain, there is no experimental data available. BMA of the uncalibrated and calibrated models can be performed, using weight factors that represent the probabilities that each model is more accurate. The weight factors are calculated using the BF obtained in the validation domain. BMA integrates results from inverse UQ and validation by simultaneously combining information from prior knowledge (in uncalibrated model), inverse UQ (in calibrated model) and quantitative validation. All major source of uncertainties can be considered in the prediction step since they have been accounted for in inverse UQ and validation.
\end{enumerate}

\section{Experimental Data}

The international OECD/NRC Boiling Water Reactor (BWR) Full-size Fine-Mesh Bundle Tests (BFBT) \cite{neykov2005nupec} benchmark was created to encourage advancement in sub-channel analysis of two-phase flow in rod bundles. In BFBT, single- and two-phase pressure losses, void fraction, and critical power tests were performed for steady-state and transient conditions. The facility is full-scale BWR assembly, with measurement performed under typical reactor power and high-pressure, high-temperature fluid conditions found in BWRs. Two types of BWR assemblies are simulated in a full length test facility, a current 8$\times$8 fuel bundle and a 8$\times$8 high burn-up bundle, where each rod is electrically heated to simulate an actual reactor fuel rod. Five test assembly configurations (types 0, 1, 2, 3 and 4) with different geometries and power profiles were utilized for the void distribution and critical power measurements. The test assembly type 0 has uniform radial and axial power distributions. Three sub-types of test bundle 0, namely 0-1, 0-2, and 0-3, were used to examine the effects of radial power distribution on the void fraction distribution by varying the number of unheated rods among them. These test assembly types/sub-types are referred to as 0011, 0021, 0031, 1071, 2081, 3091 and 4101, respectively.

\begin{figure}[!ht]
	\centering
	\includegraphics[width=0.5\textwidth]{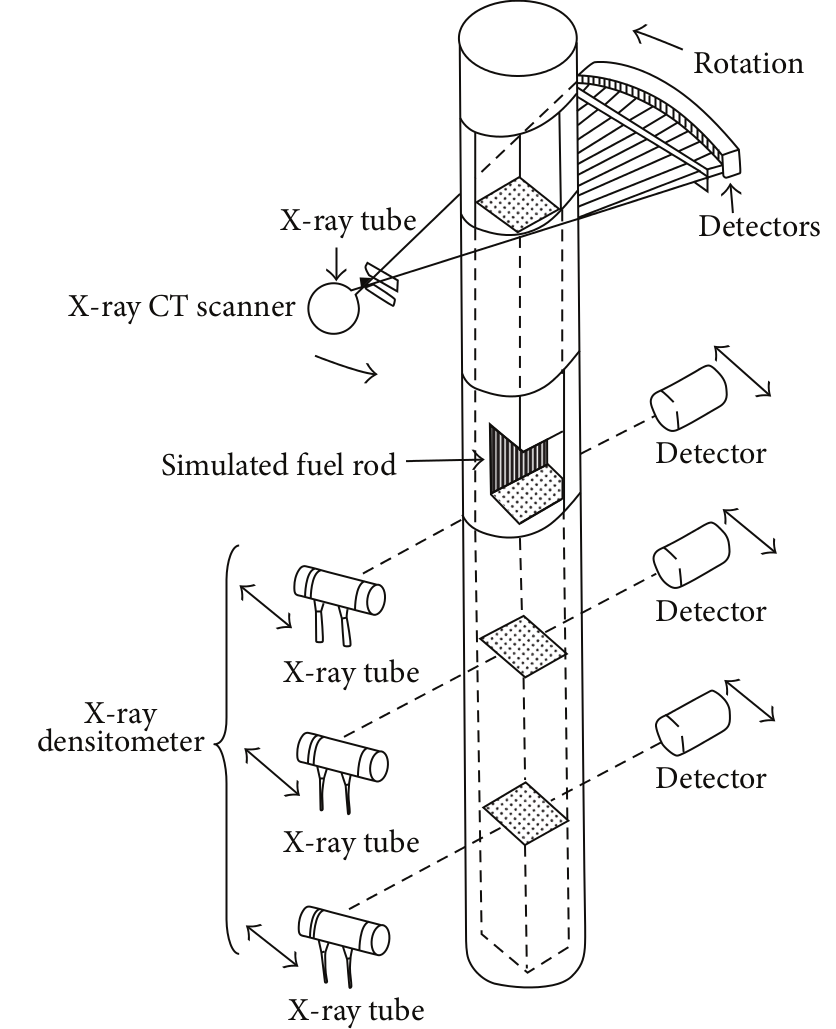}
	\caption[]{BFBT void fraction measurement structure.}
	\label{figure:fig3-BFBT-Structure}
\end{figure}

Two types of void distribution measurement systems were employed: an X-ray computer tomography (CT) scanner and an X-ray densitometer. Figure \ref{figure:fig3-BFBT-Structure} shows the void fraction measurement facility and locations. Under steady-state conditions, fine mesh void distributions were measured using the X-ray CT scanner located 50 mm above the heated length (i.e. at the assembly outlet). The X-ray densitometer measurements were performed at three different axial elevations from the bottom (i.e. 682 mm, 1706 mm and 2730 mm). For each of the four different axial locations, the cross-sectional averaged void fractions were calculated and reported in the benchmark. The steady-state void fraction data will be used in the current study, and they will be referred to from lower to upper positions as \texttt{VoidF1}, \texttt{VoidF2}, \texttt{VoidF3} and \texttt{VoidF4}, respectively.

The X-ray densitometers (used to measure \texttt{VoidF1}, \texttt{VoidF2} and \texttt{VoidF3}) can only capture the void fraction between the rod rows, therefore the measured data only shows the void fraction of a limited area of the subchannel. However, void fraction in the subchannel is not equally distributed, as pointed out in \cite{gluck2008validation}. For example, at low void fraction with bubbly flow, the void is concentrated in small bubbles close to the heat surface, while at high void fractions with slug flow, large bubbles are more likely to be located in the subchannel center. Consequently, the void fractions are under-predicted at low void fractions and over-predicted at high void fractions with the present X-ray densitometers.

To resolve this issue, data correction has been suggested in \cite{gluck2008validation}. Equation (\ref{equation:TRACE-voidF-correction1}) was proposed for assemblies 0011, 0021, 0031, 1071, 2081 and 3091, while Equation (\ref{equation:TRACE-voidF-correction2}) was suggested for assembly 4101. All the void fractions are in (\%). These equations were recommended for measured void fractions between 20\% and 90\%. Note that \texttt{VoidF4} is not corrected because it is measured by CT scanner which does not have the aforementioned limitation of an X-ray densitometer.

\begin{equation}      \label{equation:TRACE-voidF-correction1}
	\alpha_{\text{corrected}} = \frac{\alpha_{\text{measured}}}{1.231 - 0.001 \times \alpha_{\text{measured}}}
\end{equation}

\begin{equation}      \label{equation:TRACE-voidF-correction2}
	\alpha_{\text{corrected}} = \frac{\alpha_{\text{measured}}}{1.167 - 0.001 \times \alpha_{\text{measured}}}
\end{equation}

\begin{figure}[!ht]
	\centering
	\includegraphics[width=0.9\textwidth]{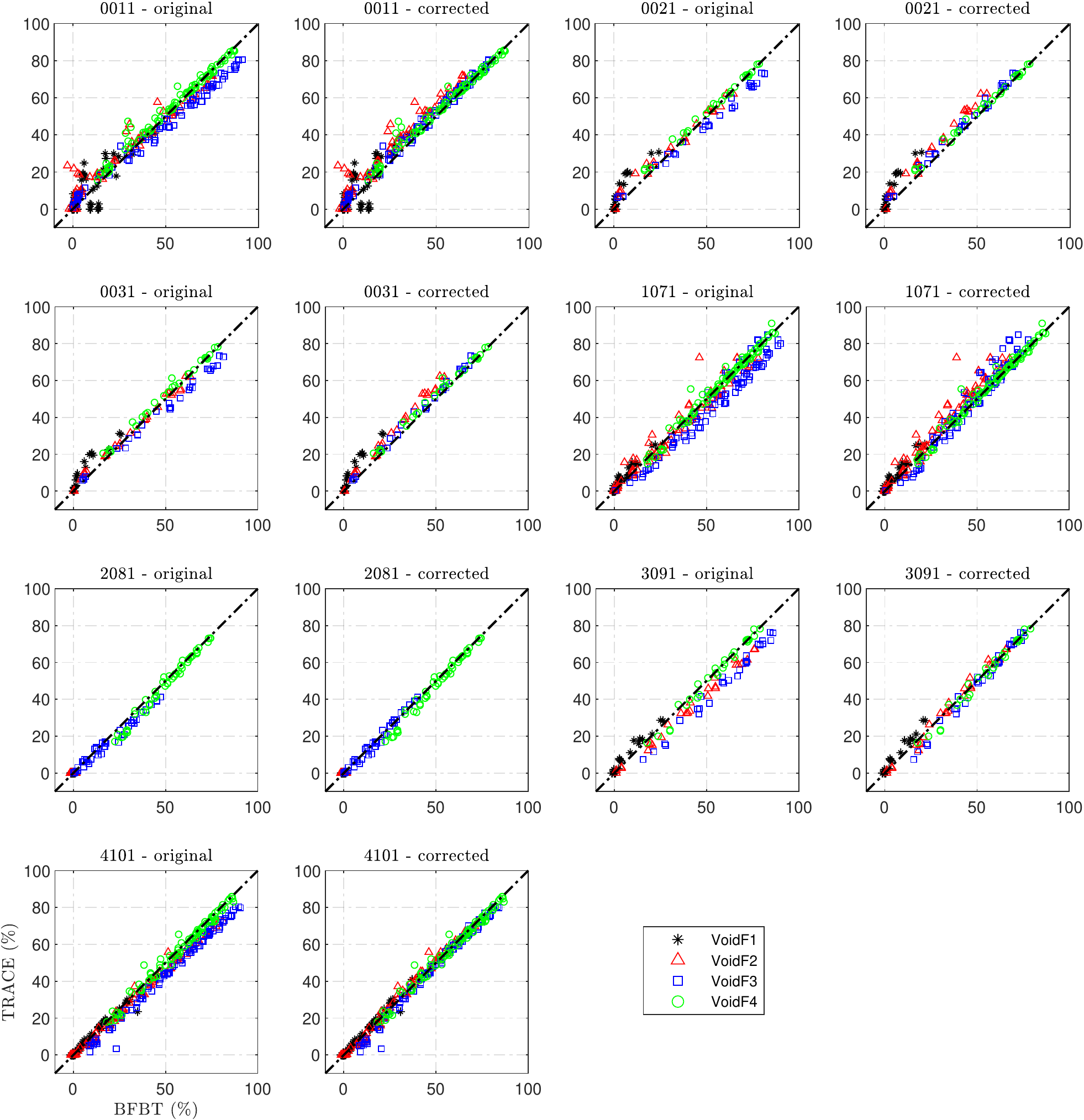}
	\caption[]{Comparison of steady-state void fraction data from BFBT benchmark and TRACE prediction at nominal values of calibration parameters.}
	\label{figure:fig3-BFBT-Data-vs-Model-Nominal}
\end{figure}

In this study, $\mathbf{y}^{\text{M}} ( \mathbf{x}, \bm{\theta} )$ is the TRACE code. The QoIs $\mathbf{y}^{\text{M}}$ are void fractions, \texttt{VoidF1}, \texttt{VoidF2}, \texttt{VoidF3} and \texttt{VoidF4}.  The design variables $\mathbf{x}$ consists of four test conditions, pressure, mass flow rate, power and inlet temperature. $\bm{\theta}$ represents the TRACE physical model parameters, which will be introduced in Section \ref{section:Inverse-UQ}. Figure \ref{figure:fig3-BFBT-Data-vs-Model-Nominal} shows a comparison of void fractions from BFBT measurements (with and without correction) and TRACE simulations (run at nominal values of $\bm{\theta}$). Before data correction, the majority of the void fractions are under predicted especially for \texttt{VoidF1}, \texttt{VoidF2} and \texttt{VoidF3}. After data correction, the data points are more concentrated close to the diagonal line, meaning that the agreement between data (BFBT) and model (TRACE) is improved.

\section{Application}

The developed methodology is applied to TRACE using the BFBT steady-state void fraction data. Significant uncertainties exist in the physical model parameters of closure laws used in nuclear reactor system thermal-hydraulics codes, which are used to describe the transfer terms in the mass, momentum and energy balance equations. TRACE version 5.0 Patch 4 \cite{USNRC2014TRACE} includes options for user access to 36 physical model parameters from the input file. The details of these parameters can be found in the TRACE user manual \cite{USNRC2014TRACE}. For forward UQ, the users can perturb these parameters by addition or multiplication according to expert judgment or user evaluation. In a previous work \cite{wu2017inverse}, we used sensitivity analysis to select 5 most significant physical model parameters relevant to the BFBT benchmark, and quantified their uncertainties with inverse UQ based on BFBT assembly 4101 steady-state void fraction data \cite{wu2018inverse-part2}. All quantified uncertainties are multiplicative factors of the nominal values. The inverse UQ results are briefly presented in Section \ref{section:Inverse-UQ}.

\subsection{Results for inverse UQ}    \label{section:Inverse-UQ}

The calibration parameters $\bm{\theta}$ consists of five uncertain physical model parameters in TRACE, including \texttt{P1008}, \texttt{P1012}, \texttt{P1022}, \texttt{P1028} and \texttt{P1029}, as shown in Table \ref{table:table2-Priors}. The nominal values are all 1.0 since they are multiplication factors. The prior ranges are chosen as $[0, 5]$ for all the parameters, which were used in design of computer experiments to build the GP metamodels. The prior ranges are chosen to be wide to reflect the ignorance of these parameters. Uniform distributions are assumed for all parameters.

\begingroup
\renewcommand{\arraystretch}{1.2}
\begin{table}[htbp]
	\footnotesize
	\captionsetup{justification=centering}
	\caption{Selected TRACE physical model parameters for inverse UQ}
	\label{table:table2-Priors}
	\centering
	\begin{tabular}{l c c c}
		\toprule
		Calibration parameters $\bm{\theta}$ (multiplication factors) & Symbol & Uniform ranges  & Nominal \\ 
		\midrule
		Single phase liquid to wall heat transfer coefficient & \texttt{P1008} & (0.0, 5.0) & 1.0 \\
		Subcooled boiling heat transfer coefficient           & \texttt{P1012} & (0.0, 5.0) & 1.0 \\
		Wall drag coefficient 								  & \texttt{P1022} & (0.0, 5.0) & 1.0 \\
		Interfacial drag (bubbly/slug Rod Bundle) coefficient & \texttt{P1028} & (0.0, 5.0) & 1.0 \\
		Interfacial drag (bubbly/slug Vessel) coefficient     & \texttt{P1029} & (0.0, 5.0) & 1.0 \\
		\bottomrule
	\end{tabular}
\end{table}
\endgroup

\begingroup
\renewcommand{\arraystretch}{1.2}
\begin{table}[htbp]
	\footnotesize
	\captionsetup{justification=centering}
	\caption{Posterior statistical moments.}
	\label{table:table3-Posteror-Moments}
	\centering
	\begin{tabular}{c c c c c}
		\toprule
		\multirow{2}{*}{Parameter} & \multicolumn{2}{c}{With model bias} & \multicolumn{2}{c}{Without model bias} \\
		\cmidrule{2-5}
		& mean values & standard deviations & mean values & standard deviations  \\ 
		\midrule
		\texttt{P1008}  &  0.6162  &  0.2113  &  1.5275 &  0.1923  \\
		\texttt{P1012}  &  1.2358  &  0.0890  &  1.0844 &  0.0592  \\
		\texttt{P1022}  &  1.4110  &  0.1833  &  0.2452 &  0.1153  \\
		\texttt{P1028}  &  1.3385  &  0.1155  &  1.4746 &  0.0414  \\
		\texttt{P1029}  &  1.2340  &  0.3453  &  0.4321 &  0.0833  \\
		\bottomrule
	\end{tabular}
\end{table}
\endgroup

Details of the inverse UQ process can be found in an earlier work \cite{wu2018inverse-part2}. Table \ref{table:table3-Posteror-Moments} shows the posterior statistical moments for $\bm{\theta}$, including the mean values and standard deviations (STDs). Two scenarios were analyzed during inverse UQ, with and without considering the model bias $\delta(\mathbf{x})$. In both cases, the statistical moments show that the posterior ranges are much narrower than the prior ranges, indicating that the knowledge in these parameters has been improved given physical observations.

\begin{figure}[htbp]
	\centering
	\includegraphics[width=0.65\textwidth]{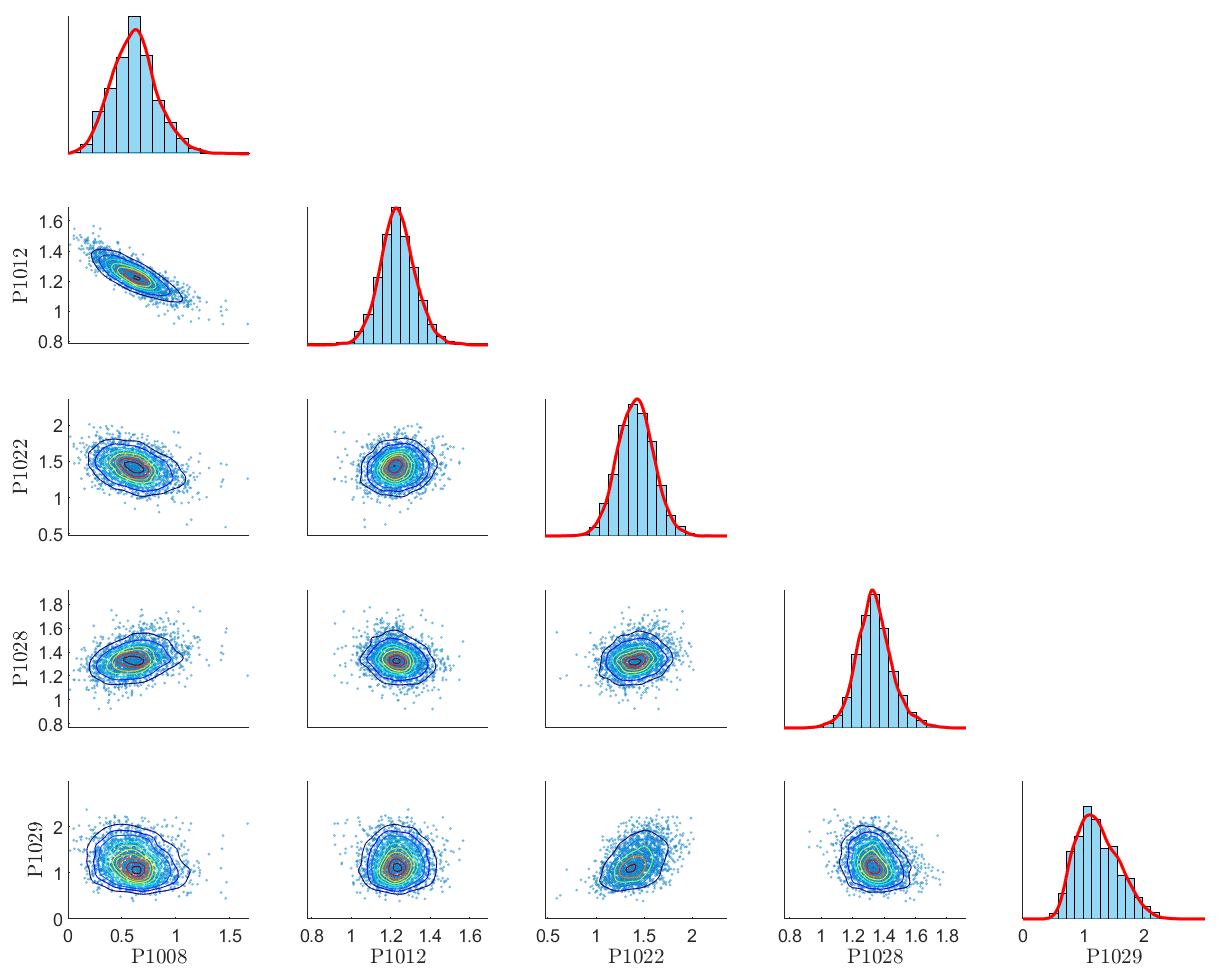}
	\caption[]{Posterior pair-wise joint (off-diagonal sub-figures) and marginal densities (diagonal sub-figures) from inverse UQ, when model discrepancy is considered. Contour plots and fitted marginal PDFs are based on MCMC samples, while scatter plots and histograms are based on copula-generated samples.}
	\label{figure:fig4-MCMC-chain-vs-Posterior-Samples-withbias}
\end{figure}

\begin{figure}[htbp]
	\centering
	\includegraphics[width=0.65\textwidth]{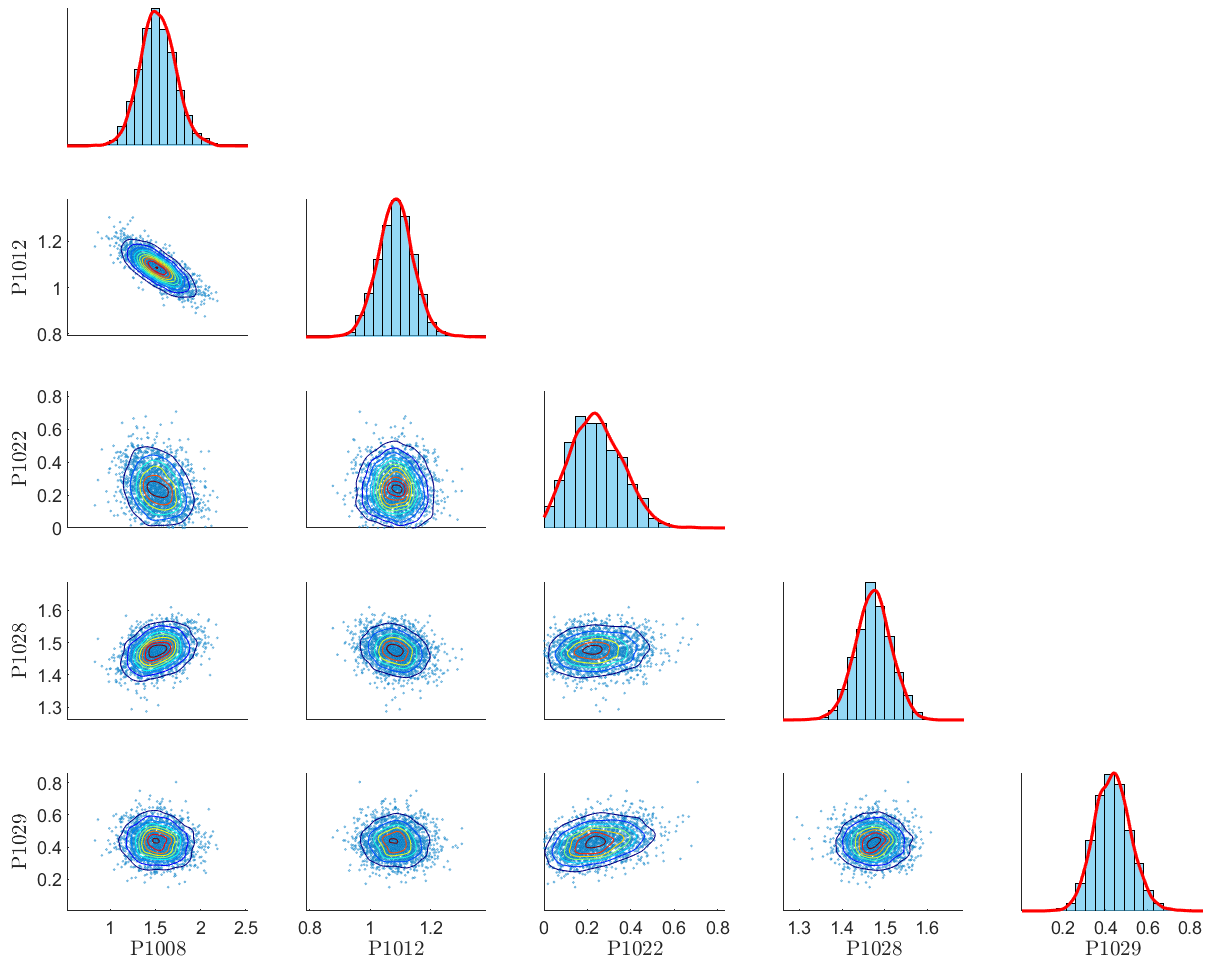}
	\caption[]{Posterior pair-wise joint (off-diagonal sub-figures) and marginal densities (diagonal sub-figures) from inverse UQ, when model discrepancy is \textbf{not} considered. Contour plots and fitted marginal PDFs are based on MCMC samples, while scatter plots and histograms are based on copula-generated samples.}
	\label{figure:fig4-MCMC-chain-vs-Posterior-Samples-nobias}
\end{figure}

Figures \ref{figure:fig4-MCMC-chain-vs-Posterior-Samples-withbias} and \ref{figure:fig4-MCMC-chain-vs-Posterior-Samples-nobias} show the posterior pair-wise joint densities and marginal densities for the five physical model parameters, with and without considering $\delta(\mathbf{x})$, respectively. Note that the contour plots and fitted marginal PDFs are obtained using the MCMC samples, while the scatter plots and histograms are new posterior samples generated to calculated the BFs (will be explained in Section \ref{section:Quantitative-Validation-with-BF}). It can be seen that the posterior samples from inverse UQ demonstrate a remarkable reduction in input uncertainties compared to their prior distributions. These density plots are also useful for identifying potential correlations between the parameters, as well as the type of marginal distribution for each parameter. For example, high negative linear correlation is observed between \texttt{P1008} and \texttt{P1012}, indicating that in future forward UQ studies, these two input parameters should be sampled jointly, not independently, so that their correlation is captured.

By comparing the inverse UQ results with and without considering $\delta(\mathbf{x})$, it can be noticed that when $\delta(\mathbf{x})$ is ignored, the posterior STDs are very small (Table \ref{table:table3-Posteror-Moments}) and pair-wise joint distributions are more concentrated (Figure \ref{figure:fig4-MCMC-chain-vs-Posterior-Samples-nobias}). This fact is preferable in the sense that more uncertainty reduction is achieved. However, it is also an indication of potential over-fitting. In Section \ref{section:Quantitative-Validation-with-BF}, models using posterior uncertainties with and without consideration of model bias will be validated using Bayesian hypothesis testing. The resulting BFs will provide a direct metric to compare these two sets of posterior distributions.

\subsection{Results for quantitative validation with Bayes factor}    \label{section:Quantitative-Validation-with-BF}

The number of tests in the seven assembly types/sub-types are listed in Table \ref{table:table4-BFBT-Number-of-Tests}. For each test, there are four void fraction measurement data available. Therefore, one BF can be calculated for each void fraction data at every test. For each assembly, we will examine the overall magnitude of the BFs averaged over all tests for \texttt{VoidF1} - \texttt{VoidF4}. The tests in each assembly will be separated randomly, with half of them used to calculate the BFs, and the other half tests used for prediction. To demonstrate the proposed methodology, we only consider assemblies 0011, 1071 and 4101, which all have over 80 tests. Assemblies 0021, 0031 and 3091 only have 28 tests, which are considered too small for both validation and prediction. Assembly 2081 has 50 tests. However, as shown in Figure \ref{figure:fig3-BFBT-Data-vs-Model-Nominal}, \texttt{VoidF1} and \texttt{VoidF2} are all zeros in both experiments and TRACE simulations. Therefore, assembly 2081 is also not considered in this study.

\begin{table}[htbp]
	\footnotesize
	\captionsetup{justification=centering}
	\caption{Number of tests available in BFBT's seven assembly types/sub-types.}
	\label{table:table4-BFBT-Number-of-Tests}
	\centering
	\begin{tabular}{cccccccc}
		\toprule
		Assemblies  &  0011  &  0021  &  0031  &  1071  &  2081  &  3091  &  4101  \\ 
		\midrule
		Number of tests  &  83  &  28  &  28  &  86  &  50  &  28  &  86  \\
		\bottomrule
	\end{tabular}
\end{table}

The tests in assembly 4101 were already used for inverse UQ in the earlier work \cite{wu2018inverse-part2}. Ideally, data that has been used for calibration should not be used again for validation and prediction. However, in this paper we still consider assembly 4101 data for the purpose of ``verification'' of the proposed methodology. Assemblies 1011 and 1071 provide completely ``blind'' data that has not been used during inverse UQ. Application of the quantified posterior uncertainties based on assembly 4101 tests, to the tests in assemblies 1011 and 1071, is essentially an extrapolation step. Such extrapolation can demonstrate whether the posterior uncertainties can improve TRACE predictions for new experiments.

The tests in assemblies 1011, 1071 and 4101 are randomly partitioned to validation sets and prediction sets. Based on the validation tests, we will calculate the BFs for TRACE using posterior uncertainties with and without considering the model bias $\delta(\mathbf{x})$. The prediction tests will be used in Section \ref{section:Prediction-by-Integration} to perform BMA, by averaging the predictions based on prior and posterior distributions using the BFs. The BFs are calculated by comparing TRACE simulation using posterior and prior distributions. Monte Carlo integration is used to evaluate the integrals in Equation (\ref{equation:BF2}) by generating a large number of random samples for $\bm{\theta}$. These samples are generated based on the MCMC samples using copula. Copulas are functions that can describe dependencies among multiple correlated variables. The application of copula can generate arbitrary number of random samples that share the same joint and marginal posterior distributions with the MCMC samples. The generated posterior samples are presented as scatter plots and histograms in Figures \ref{figure:fig4-MCMC-chain-vs-Posterior-Samples-withbias} and \ref{figure:fig4-MCMC-chain-vs-Posterior-Samples-nobias}. Validation needs to calculate the TRACE model output for each sample, which is very computationally expansive. A GP metamodel is used to reduce the cost.

\begingroup
\renewcommand{\arraystretch}{1.2}
\begin{table}[htbp]
	\footnotesize
	\captionsetup{justification=centering}
	\caption{BFs for validation with steady-state void fraction data from assemblies 0011, 1071 and 4101.}
	\label{table:table4-Bayes-Factors}
	\centering
	\begin{tabular}{c l c c c c}
		\toprule
		Assembly  &  Models  & \texttt{VoidF1}  &  \texttt{VoidF2}  &  \texttt{VoidF3}  &  \texttt{VoidF4}  \\ 
		\midrule
		\multirow{2}{*}{0011}  
		& Without model bias  &  1.1663  &  0.3296  &  0.4354  &  1.1143  \\
		& With model bias     &  1.9140  &  0.4518  &  1.0746  &  1.2751  \\
		\midrule
		\multirow{2}{*}{1071}  
		& Without model bias  &  1.0718  &  0.5692  &  1.1366  &  1.3501  \\
		& With model bias     &  1.4447  &  0.6455  &  1.2576  &  1.5493  \\
		\midrule
		\multirow{2}{*}{4101}  
		& Without model bias  &  4.2967  &  1.6851  &  2.7487  &  3.4627  \\
		& With model bias     &  5.0553  &  1.7191  &  2.1205  &  1.3720  \\
		\bottomrule
	\end{tabular}
\end{table}
\endgroup

\begin{figure}[htbp]
	\centering
	\includegraphics[width=0.99\textwidth]{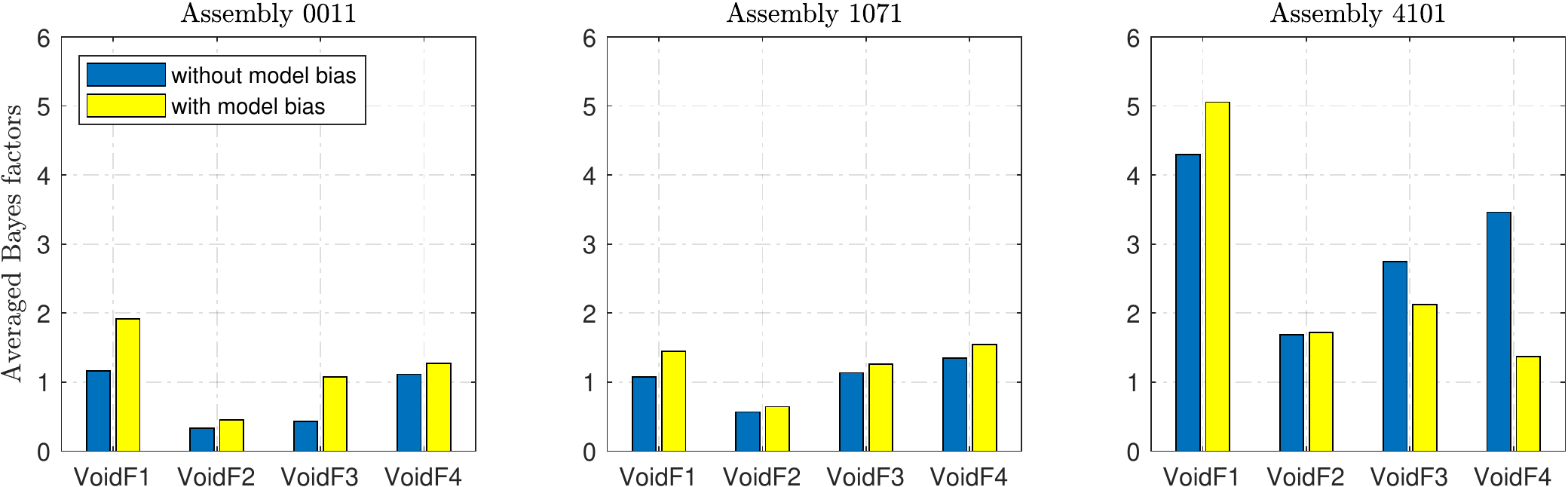}
	\caption[]{Calculated BF values, averaged over different tests in assemblies 0011, 1071 and 4101.}
	\label{figure:fig5-BF-Results-Mean}
\end{figure}

Table \ref{table:table4-Bayes-Factors} presents the BFs for posterior distributions with and without considering the model bias $\delta(\mathbf{x})$ in Table \ref{table:table3-Posteror-Moments}, by comparing with the prior distributions in Table \ref{table:table2-Priors}. For each of 0011, 1071 and 4101, the values in Table \ref{table:table4-Bayes-Factors} are averaged over all the validation tests for \texttt{VoidF1} - \texttt{VoidF4}. The BFs are visualized in Figure \ref{figure:fig5-BF-Results-Mean}. It can be observed that:
\begin{enumerate}[label=(\arabic*)]
	\setlength{\itemsep}{0.1pt}
	\item For assembly 4101, the BFs are larger than the other two assemblies. Since the same data was used for inverse UQ, the posterior distributions are expected to lead to better agreement with data and higher BFs. Furthermore, posterior model without considering the model bias has higher BFs for \texttt{VoidF3} and \texttt{VoidF4}, which is a sign for over-fitting. Such results are not found in \texttt{VoidF1} and \texttt{VoidF2}, because these void fractions are generally much smaller or even close to 0, making over-fitted posterior distributions that have narrow ranges unable to re-produce the data.
	
	\item For assemblies 0011 and 1071, posterior model that considered the model bias consistently have high BFs than the posterior model without model bias. However, the BFs are much loser to 1.0 or even smaller than 1. Recall that a BF value of 1.0 means that the data have equal support for the posterior and prior models. In other words, when the posterior distributions obtained using assembly 4101 tests is extrapolated to new tests, the calibrated model is less likely to reproduce the new data. But still, the calibrated models have overall better performance than the prior model, because the BFs are greater than 1.0 except for \texttt{VoidF2}.
\end{enumerate}

The quantitative validation results indicate that, using the posterior distributions from inverse UQ, TRACE's predictive capability has been improved in most tests. However, the improvement is not significant. The main reason is that the agreement between TRACE predictions and the BFBT void fraction data is already very good before inverse UQ, as shown in Figure \ref{figure:fig3-BFBT-Data-vs-Model-Nominal}. Furthermore, the benefit of considering model bias in inverse UQ in this application can be verified, thought not substantial, due to the fact that TRACE predicts well before inverse UQ so the model bias is relatively small.

\subsection{Results for prediction by integrating inverse UQ and quantitative validation}     \label{section:Prediction-by-Integration}

To show that the proposed methodology can improve TRACE predictive capability, five models will be compared in this section, as listed in Table \ref{table:table6-BMA-Models}. Model A represents uncalibrated TRACE using the prior distributions listed in Table \ref{table:table2-Priors}. Model B represents calibrated TRACE using the posterior distributions that were obtained without consideration of $\delta(\mathbf{x})$. Model C is similar to model B, but uses the posterior distributions with considering $\delta(\mathbf{x})$. Model D denotes a model that performs BMA of A and B, using the BFs calculated from these two models. Similarly, model E is the BMA of A and C.

\begingroup
\renewcommand{\arraystretch}{1.3}
\begin{table}[htbp]
	\footnotesize
	\captionsetup{justification=centering}
	\caption{Models that are used in the prediction domain.}
	\label{table:table6-BMA-Models}
	\centering
	\begin{tabular}{cc}
		\toprule
		Models  &  Details  \\ 
		\midrule
		A  &  TRACE using $\bm{\theta}^{\text{Prior}}$  \\
		B  &  TRACE using $\bm{\theta}^{\text{Posterior}}$ without model bias  \\
		C  &  TRACE using $\bm{\theta}^{\text{Posterior}}$ with model bias  \\
		D  &  BMA of models A and B  \\
		E  &  BMA of models A and C  \\
		\bottomrule
	\end{tabular}
\end{table}
\endgroup

The BFs are used to calculate the weight factors for BMA according to Equation (\ref{equation:Prediction1-Weight-Factor}). The weight factors that correspond to the BFs in Table \ref{table:table4-Bayes-Factors} are presented in Table \ref{table:table6-BMA-Weight-Factors}. The weights for posterior models are larger than 0.5 when the BFs are greater than 1. Larger BFs mean the corresponding posterior distributions have more support from the validation data, leading to larger weights for the calibrated model using these posterior distributions. Equation (\ref{equation:Prediction2-Weighted-Average}) is used to perform the model averaging.

\begingroup
\renewcommand{\arraystretch}{1.2}
\begin{table}[htbp]
	\footnotesize
	\captionsetup{justification=centering}
	\caption{Weight factors for BMA, calculated based on the BFs.}
	\label{table:table6-BMA-Weight-Factors}
	\centering
	\begin{tabular}{c c c c c c c c c c}
		\toprule
		\multirow{2}{*}{Assembly}  & \multirow{2}{*}{Model} & \multicolumn{4}{c}{weights for model A} & \multicolumn{4}{c}{wights for model B/C}  \\
		\cmidrule{3-10}
		& & \texttt{VoidF1}  &  \texttt{VoidF2}  &  \texttt{VoidF3}  &  \texttt{VoidF4}  & \texttt{VoidF1}  &  \texttt{VoidF2}  &  \texttt{VoidF3}  &  \texttt{VoidF4}  \\ 
		\midrule
		\multirow{2}{*}{0011}  
		&  D  &  0.4616  &  0.7521  &  0.6967  &  0.4730  &  0.5384  &  0.2479  &  0.3033  &  0.5270  \\
		&  E  &  0.3432  &  0.6888  &  0.4820  &  0.4395  &  0.6568  &  0.3112  &  0.5180  &  0.5605  \\
		\midrule
		\multirow{2}{*}{1071}  
		&  D  &  0.4827  &  0.6373  &  0.4680  &  0.4255  &  0.5173  &  0.3627  &  0.5320  &  0.5745  \\
		&  E  &  0.4090  &  0.6077  &  0.4430  &  0.3923  &  0.5910  &  0.3923  &  0.5570  &  0.6077  \\
		\midrule
		\multirow{2}{*}{4101}  
		&  D  &  0.1888  &  0.3724  &  0.2668  &  0.2241  &  0.8112  &  0.6276  &  0.7332  &  0.7759  \\
		&  E  &  0.1651  &  0.3678  &  0.3205  &  0.4216  &  0.8349  &  0.6322  &  0.6795  &  0.5784  \\
		\bottomrule
	\end{tabular}
\end{table}
\endgroup

\begin{figure}[htbp]
	\centering
	\includegraphics[width=0.99\textwidth]{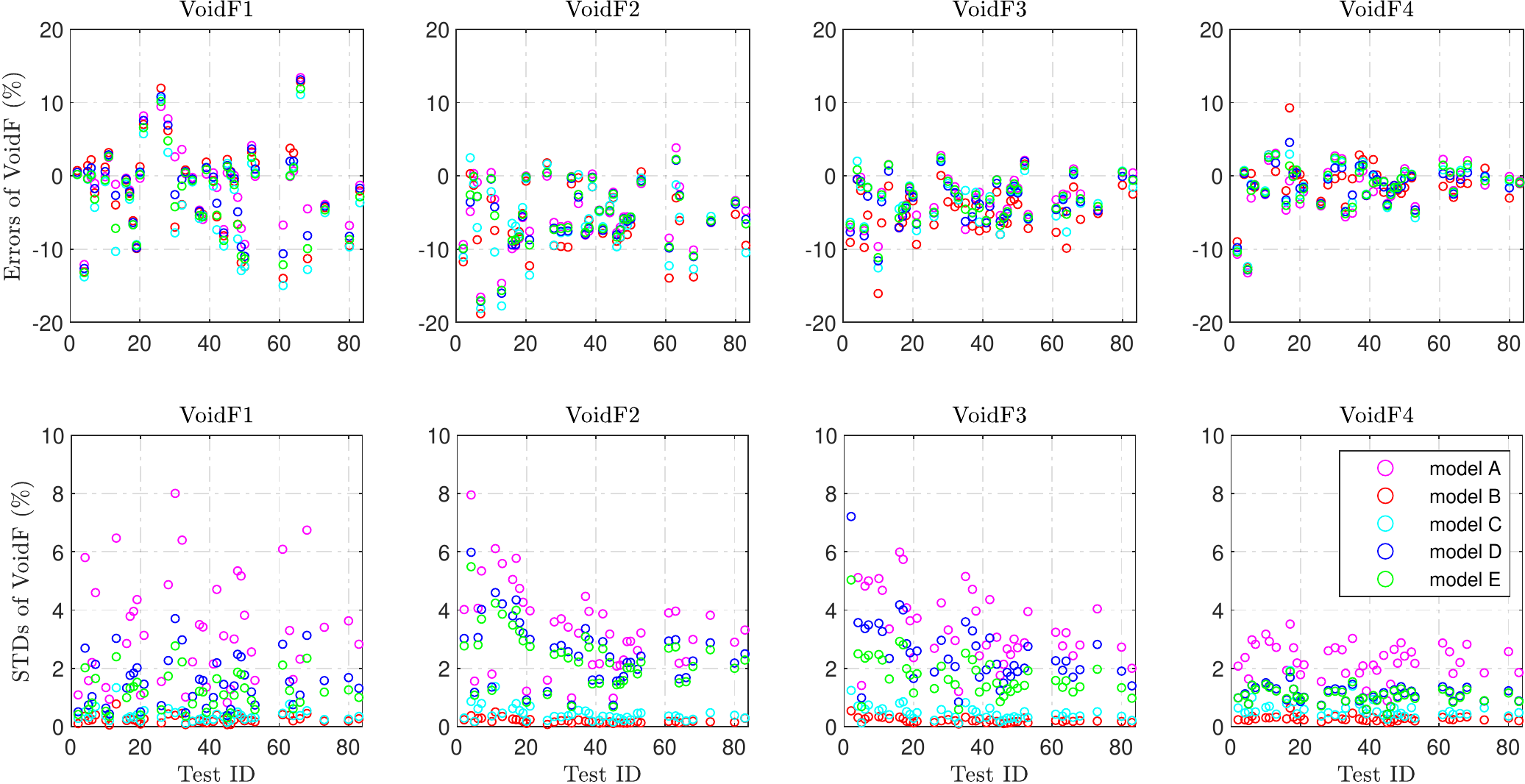}
	\caption[]{Errors in void fraction mean predictions, and the void fractions STDs for assembly 0011.}
	\label{figure:fig6-BMA-Results-0011}
\end{figure}

\begin{figure}[htbp]
	\centering
	\includegraphics[width=0.99\textwidth]{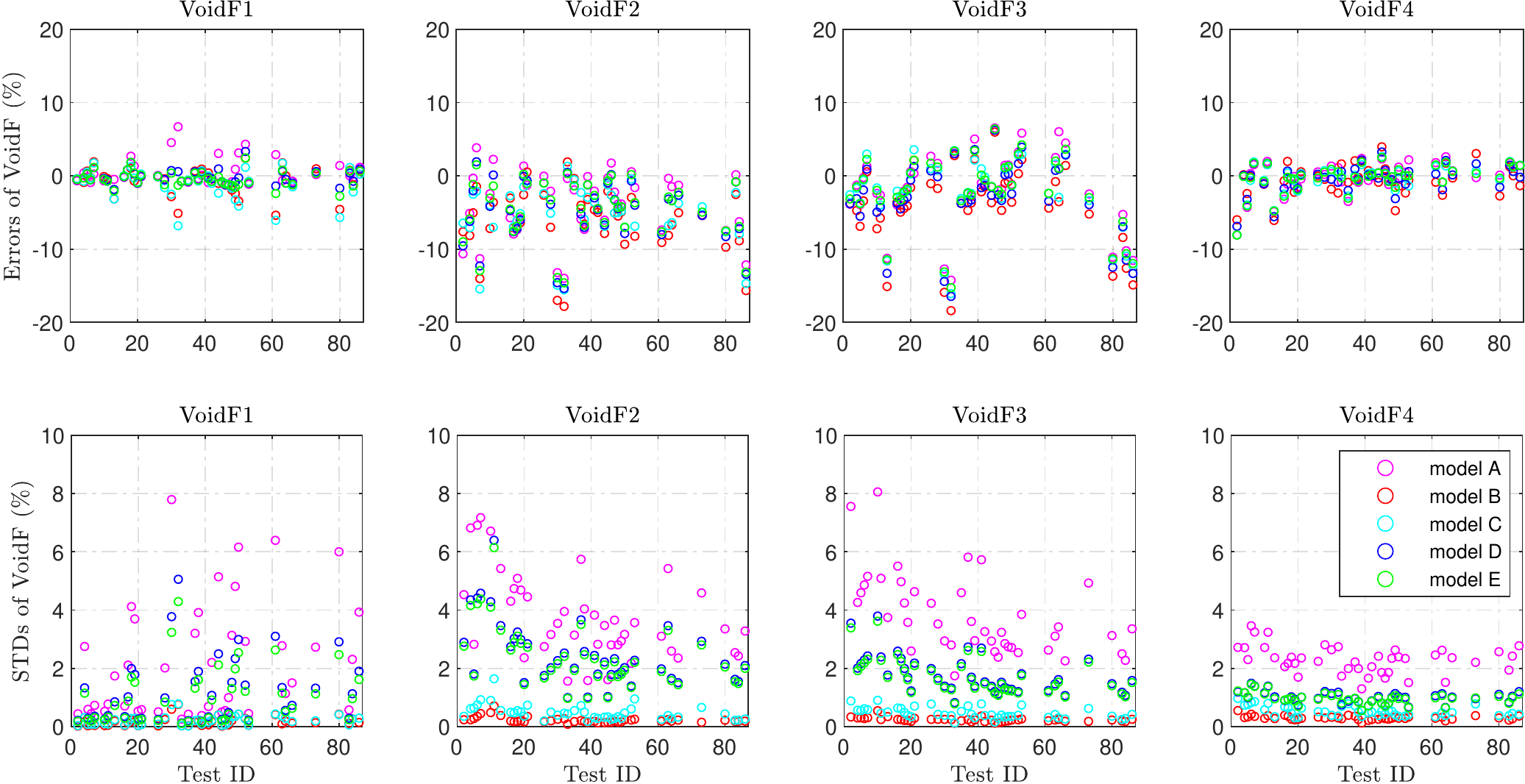}
	\caption[]{Errors in void fraction mean predictions, and the void fractions STDs for assembly 1071.}
	\label{figure:fig6-BMA-Results-1071}
\end{figure}

\begin{figure}[htbp]
	\centering
	\includegraphics[width=0.99\textwidth]{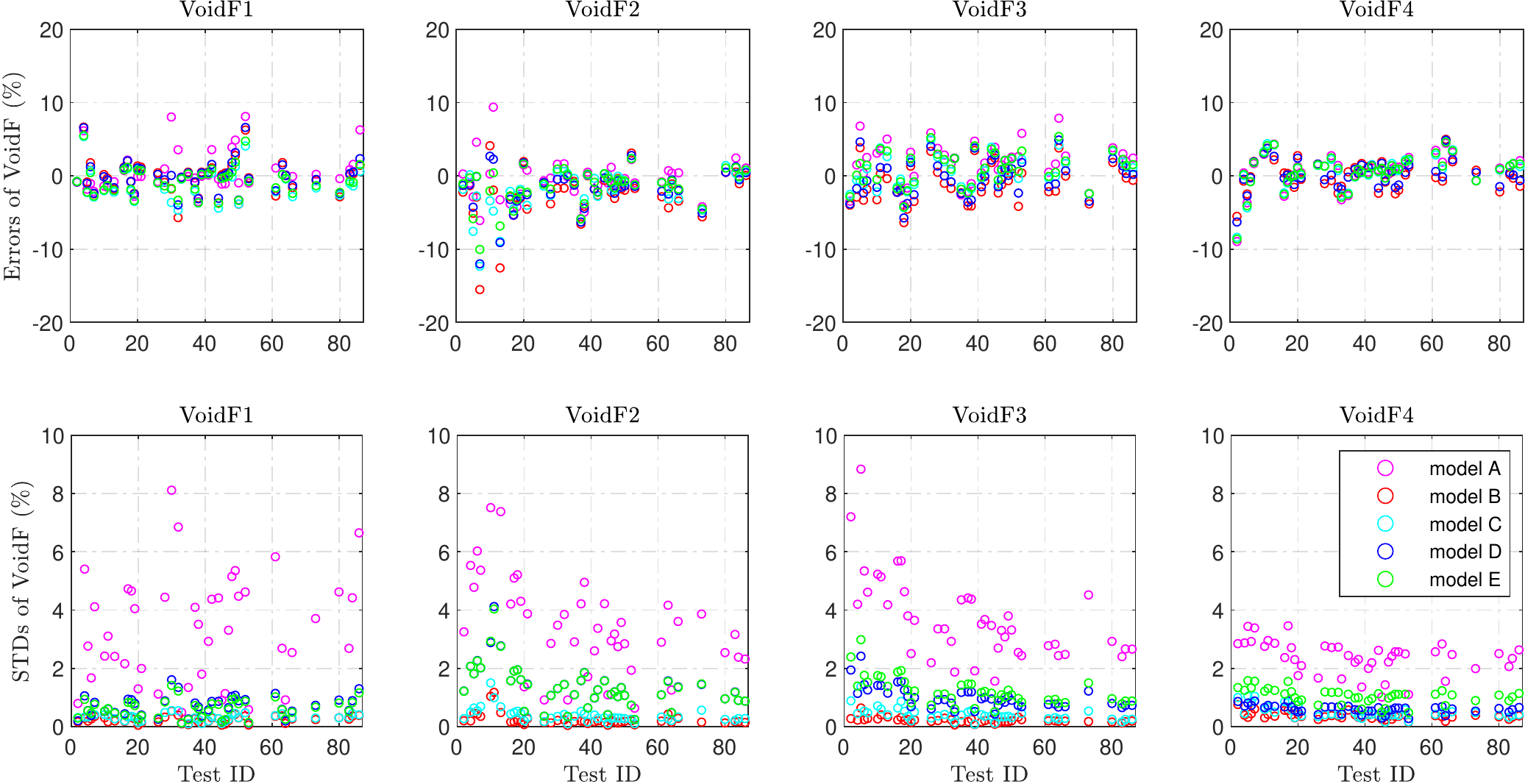}
	\caption[]{Errors in void fraction mean predictions, and the void fractions STDs for assembly 4101.}
	\label{figure:fig6-BMA-Results-4101}
\end{figure}

To obtain the predictions from models A-E, 2,000 samples are generated based on the prior and posterior distributions (using copula). These samples are evaluated again using an accurate GP metamodel. The void fraction predictions for models A-C can be directly obtained, while the results for models D and E are calculated with BMA, using the weight factors in Table \ref{table:table6-BMA-Weight-Factors}, for each of \texttt{VoidF1} - \texttt{VoidF4}. The mean values and STDs of these 2,000 void fraction predictions are then calculated. Figures \ref{figure:fig6-BMA-Results-0011} - \ref{figure:fig6-BMA-Results-4101} shows the results for BMA applied to the prediction tests in assemblies 0011, 1071 and 4101, respectively. In each assembly, BFBT void fraction data is compared to the mean values of the void fraction predictions from models A-E, and their differences for each test in shown in the top row. In the bottom row, the STDs of the void fraction predictions are presented, which represent the uncertainties in the predictions.

For the STDs, model A produces the largest uncertainties, which is expected since the prior distributions are wide. Model B leads to the smallest uncertainties, because it uses posteriors obtained without considering the model bias, which are more concentrated and has been proven to be over-fitted to the inverse UQ data. Model C also results in very small QoI uncertainties. Models from BMA (D/E) have moderate STDs. In assembly 4101, the QoI uncertainties from models D/E are close to those from model B/C, because the large BF values cause greater weights towards the calibration models. As a comparison, in assemblies 0011 and 1071, models D/E have QoI uncertainties that are between model A and models B/C. Such moderate QoI uncertainties are desirable, as they represent an uncertainty reduction from model A that uses wide prior ranges, while still maintain slight conservatism by not completely relying on posterior distributions.

\begin{figure}[htbp]
	\centering
	\includegraphics[width=0.99\textwidth]{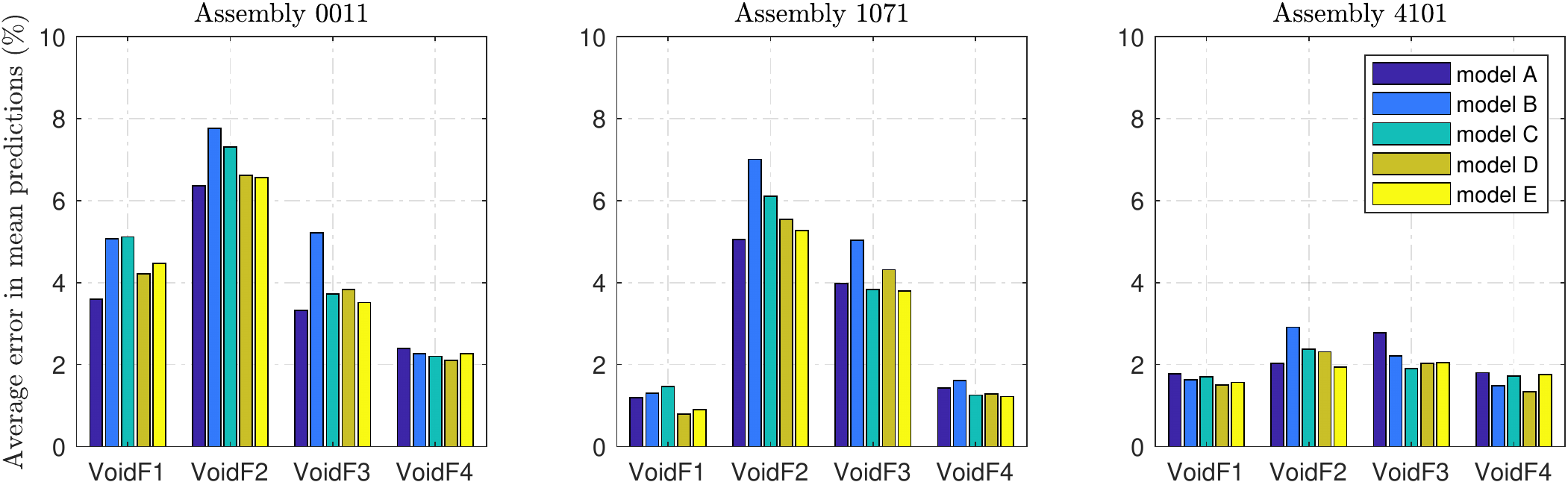}
	\caption[]{Errors in void fraction mean predictions, averaged over different tests in assemblies 0011, 1071 and 4101, respectively.}
	\label{figure:fig7-BMA-Results-Averaged-Error}
\end{figure}

For the errors in void fraction mean predictions, no substantial differences can be observed for models A-E. They all agree in a similar way with BFBT experimental data. To better compare these models, the errors in void fraction mean predictions, shown in the top rows of Figures \ref{figure:fig6-BMA-Results-0011} - \ref{figure:fig6-BMA-Results-4101}, are averaged over different tests. The results are presented in Table \ref{table:table6-BMA-Results-Averaged-Error} and visualized in Figure \ref{figure:fig7-BMA-Results-Averaged-Error}. It can be observed that:
\begin{enumerate}[label=(\arabic*)]
	\setlength{\itemsep}{0.1pt}
	\item For assemblies 0011 and 1071, model B generally have the largest averaged errors due to over-fitting. Models D/E have better performance than model B/C after averaging them with model A. But the improvements of models D/E over model A is very small. For \texttt{VoidF1} - \texttt{VoidF3} in assembly 0011, and \texttt{VoidF2} in assembly 1071, models D/E have larger averaged errors than model A.
	
	\item For assembly 4101, models B-E generally have slightly better performance than model A except for \texttt{VoidF2}, which is expected since the data in assembly 4101 has been used for inverse UQ. The averaged errors are also smaller than the other two assemblies.
\end{enumerate}

\begingroup
\renewcommand{\arraystretch}{1.2}
\begin{table}[htbp]
	\footnotesize
	\captionsetup{justification=centering}
	\caption{Errors in void fraction mean predictions, averaged over different tests.}
	\label{table:table6-BMA-Results-Averaged-Error}
	\centering
	\begin{tabular}{c c c c c c}
		\toprule
		Assembly  &  Model  &  \texttt{VoidF1}  &  \texttt{VoidF2}  &  \texttt{VoidF3}  &  \texttt{VoidF4}  \\ 
		\midrule
		\multirow{5}{*}{0011}  
		&  A  &  3.5937  &  6.3550  &  3.3261  &  2.3964  \\
		&  B  &  5.0648  &  7.7689  &  5.2151  &  2.2656  \\
		&  C  &  5.1164  &  7.3102  &  3.7229  &  2.2019  \\
		&  D  &  4.2151  &  6.6222  &  3.8288  &  2.0961  \\
		&  E  &  4.4741  &  6.5602  &  3.5117  &  2.2608  \\
		\midrule
		\multirow{5}{*}{1071}  
		&  A  &  1.1865  &  5.0487  &  3.9790  &  1.4255  \\
		&  B  &  1.3023  &  7.0051  &  5.0343  &  1.6098  \\
		&  C  &  1.4617  &  6.1057  &  3.8287  &  1.2548  \\
		&  D  &  0.7877  &  5.5483  &  4.3141  &  1.2780  \\
		&  E  &  0.9008  &  5.2642  &  3.7992  &  1.2189  \\
		\midrule
		\multirow{5}{*}{4101}  
		&  A  &  1.7724  &  2.0249  &  2.7713  &  1.8043  \\
		&  B  &  1.6319  &  2.9104  &  2.2119  &  1.4849  \\
		&  C  &  1.6978  &  2.3751  &  1.8978  &  1.7174  \\
		&  D  &  1.4988  &  2.3098  &  2.0313  &  1.3398  \\
		&  E  &  1.5651  &  1.9450  &  2.0448  &  1.7533  \\
		\bottomrule
	\end{tabular}
\end{table}
\endgroup

\subsection{Summary and discussion}

Bayesian hypothesis testing has rarely been used to perform quantitative validation in the nuclear engineering area. The results in Section \ref{section:Quantitative-Validation-with-BF} have shown that the calculated BFs can serve as a rigorous metric to quantitatively assess the model accuracy. It was shown that when inverse UQ is performed without considering the model bias, the results are highly possible to be over-fitted to the chosen experimental data. In Section \ref{section:Prediction-by-Integration}, the calibrated models and BMA models show small (or no) improvements over the uncalibrated model that uses the prior, when the posterior uncertainties are extrapolated to new experimental tests. However, even though the averaged errors in void fraction mean predictions from BMA models are generally very close to the uncalibrated model, the resulting QoI uncertainties are usually reduced substantially. Based on these results, BMA has been shown to produce improved forward UQ results by combining prior knowledge, inverse UQ and validation results through Bayesian hypothesis testing.

\section{Conclusions}

In this paper, we proposed a framework that combines experimentation and M\&S into a unified approach to improve the predictive capabilities of computer models. This framework integrates inverse UQ (calibration) and quantitative validation to improve prediction, while accounting for all major sources of quantifiable uncertainties in M\&S, i.e., uncertainties from parameters, experiment, model and code. A quantitative validation metric called Bayes factor was formulated through Bayesian hypothesis testing, and was used to calculate the weight factors in a Bayesian model averaging process. It can improve forward UQ results by combining information from prior knowledge (mainly based on expert opinion), calibration (inverse UQ) and quantitative validation. Even though the averaged errors in void fraction mean predictions from such integration are generally very close to the prior uncalibrated model, the resulting response uncertainties are reduced substantially. This framework serves as an initial step towards providing a feasible solution to the ANS Nuclear Grand Challenge on ``Simulation \& Experimentation'' and can enable reduced reliance on expensive measurement data.

\bibliography{./Journal_Integrated_Framework.bib}

\end{document}